\documentclass{elsarticle}

\usepackage[english]{babel}

\usepackage[letterpaper,top=2cm,bottom=2cm,left=3cm,right=3cm,marginparwidth=1.75cm]{geometry}

\usepackage{amsmath,amssymb}
\usepackage{graphicx}
\usepackage[colorlinks=true, allcolors=blue]{hyperref}
\usepackage{bm}
\usepackage{algorithm}
\usepackage{algpseudocode}
\usepackage{tabularx}
\usepackage{booktabs}
\usepackage{siunitx}
\DeclareSIUnit\normalcubicmeter{Nm^3}
\DeclareSIUnit\normalcubicmeterpersec{Nm^3/s}
\DeclareSIUnit\terawatthoursperyear{TWh/a}
\DeclareSIUnit\bar{bar}

\usepackage[markup=underlined]{changes}
\definechangesauthor[name={Ulrike}, color=blue]{UH}

\newcommand{\R}{\mathbb{R}}
\newcommand{\N}{\mathbb{N}}

\newcommand{\mrm}[1]{\mathrm{#1}}
\newcommand{\mc}[1]{\mathcal{#1}}
\newcommand{\cs}{c_\mrm{s}}

\newtheorem{defi}{Definition}[section]

\newtheorem{problem}[defi]{Problem}

\begin{document}

\begin{frontmatter}

\title{Quantifying Linepack Storage Potential of Hydrogen Networks}
\author[ieg,btu]{Ulrike Herrmann}
\author[pik]{Hans Würfel}
\author[voi]{Albertus Johannes Malan\fnref{fn1}}
\author[ieg]{Anton Plietzsch}
\author[ieg,btu]{Johannes Schiffer}

\affiliation[ieg]{organization={Fraunhofer IEG, Fraunhofer Research Institution for Energy
  Infrastructures and Geotechnologies IEG},
  postcode={03046}, city={Cottbus}, country={Germany}}

\affiliation[pik]{organization={Potsdam Institute for Climate Impact Research (PIK)},
  postcode={14473}, city={Potsdam}, country={Germany}}

\affiliation[voi]{organization={Voith Hydro Holding GmbH \& Co. KG},
  postcode={89522}, city={Heidenheim}, country={Germany}}

\affiliation[btu]{organization={Brandenburg University of Technology
  Cottbus-Senftenberg},
  postcode={03046}, city={Cottbus}, country={Germany}}

\fntext[fn1]{This work was conducted independently and does not represent the views or work of Voith Hydro Holding GmbH \& Co. KG.}

\begin{abstract}
Future hydrogen systems depend on large-scale storage to balance supply and demand. Hydrogen networks can store energy in the pipelines themselves, but the potential of this linepack storage has not been quantified comparably to dedicated storage facilities. We transfer metrics established for conventional storage, such as volume capacity and full-cycle equivalents, to hydrogen networks. To evaluate them, we formulate an optimization problem maximizing linepack storage utilization, subject to non-convex PDE-based constraints describing transient gas flow. We solve the optimization problem over a full-year horizon to capture seasonal variations and to assess the interaction with cavern storage. For computational tractability, we propose a simulation-informed sequential convex programming approach with adaptive time stepping that reduces problem size by up to \qty{98}{\percent} at accuracy comparable to fixed-resolution methods. For the GasLib-40 topology with German scenario data, the network provides a volume capacity of \qty{24e6}{\normalcubicmeter} at \num{28} full-cycle equivalents per year.
\end{abstract}


\end{frontmatter}

\section{Introduction}\label{sec:intro}

\subsection{Motivation}

Future hydrogen systems will require significant storage capacity to bridge temporal imbalances between supply and demand across time scales from hours to seasons~\cite{Lux.2022, Neumann.2023, Kondziella.2023, Frischmuth.2024}.
On a seasonal scale, storage demand arises primarily from hydrogen-fired power generation during periods of low renewable availability. On shorter, intraday time scales, storage is required to buffer the volatile output of renewable-powered electrolyzers. In most energy system models, the entire hydrogen storage demand is assigned to underground storage facilities such as salt caverns, which represent the most cost-effective solution for large-scale storage~\cite{Elberry.2021, Papadias.2021}. However, these facilities are limited in their operational flexibility: frequent and intensive charging and discharging cycles can significantly reduce their service lifetime~\cite{Li.2023, RameshKumar.2023}. Complementary short-term storage options are therefore likely to be required in real-world systems, 
yet such alternatives are either not represented or only very simplistically modeled in most energy system frameworks~\cite{Lux.2022, Neumann.2023, Kondziella.2023, Frischmuth.2024}.

Potential short-term hydrogen storage options include dedicated underground pipe storage and the inherent linepack storage capacity of the hydrogen transport 
network itself. Pipe storage, well established in the natural gas sector, is incorporated into some energy system models but is rarely selected in 
optimization due to its relatively high cost~\cite{Papadias.2021}. An alternative that has received little attention so far is linepack storage, 
i.e., the temporary storage of hydrogen within the transport pipelines by varying the network pressure level~\cite{Frischmuth.2024}. Unlike dedicated 
storage facilities, linepack requires no additional infrastructure and is inherently available in any pipeline network.

However, energy system models are fundamentally unsuited to quantify linepack storage potential. These models typically represent gas networks using 
steady-state or linear flow balances that assume an instantaneous equilibrium between supply and demand at each time step~\cite{Lux.2022, Neumann.2023, Kondziella.2023, Frischmuth.2024}. 
Under such assumptions, no mass can accumulate within the network, and the concept of linepack — which arises precisely from transient pressure and mass 
flow variations — cannot be captured. A physically consistent assessment of linepack storage therefore requires a transient fluid-mechanical network 
model that resolves the temporal evolution of pressure and flow throughout the pipeline system.

No detailed analysis of the technical potential of linepack storage in hydrogen networks has been conducted to date. We 
therefore address the following research question: \textit{How can the maximum technically feasible potential of linepack storage in a hydrogen 
network be quantified?} 
In answering this question, we focus exclusively on the technical storage potential as determined by network topology, operational constraints, and transport scenarios. Economic viability and regulatory considerations are beyond the scope of this work.

\subsection{Existing Literature}\label{sec:literature}
The existing literature addressing linepack storage can broadly be divided into three strands: (i)~studies quantifying the storage potential of individual pipelines, (ii) research on the transient operational optimization of gas networks, in which linepack storage is exploited implicitly,
and (iii) a few works that explicitly maximize the linepack storage potential of gas networks.

The first strand includes~\cite{Sargent.2025}, which compares the linepack storage potential of natural gas and hydrogen for a single pipeline. However, the assessment is limited to operation at the design point of the pipeline. Its applicability to entire networks under varying load conditions is therefore limited. Moreover, the proposed storage metric does not allow for a direct comparison with other types of storage. 
In~\cite{Jottrand.2025}, a single pipeline is used as a buffer for the mismatch between supply and
demand in a simulation with full-year horizon. The annual supply--demand mismatch is scaled down until the operational limits of the pipeline are satisfied, and the remaining imbalance is absorbed by unspecified buffer
storage. Neither the size of this external buffer storage nor a metric for the pipeline storage potential are reported.

The second strand comprises a substantial body of research on the operational optimization of gas networks, addressing
Optimal Gas Flow (OGF) (e.g.,~\cite{Zlotnik.2019, OMalley.2021, Shi.2026}) or Optimal Power and Gas Flow (OPGF)
problems (e.g.,~\cite{Mhanna.2022, Raheli.2025, Schwele.2019, Fan.2025}) using transient gas network models. In these studies, linepack storage is utilized implicitly, without being the main focus. Schwele et
al.~\cite{Schwele.2019}, for instance, minimize production costs in the unit dispatch of a coupled
electricity--natural gas system and quantify the resulting cost savings enabled by linepack flexibility, but
they do not derive a storage metric as such\footnote{Schwele et al.~\cite{Schwele.2019} refer to their model as a steady-state model. However, they combine a steady-state momentum balance with a dynamic mass balance. This approach, which is also used in~\cite{Burlacu.2019}, is known as the friction-dominated model~\cite{Burlacu.2019, Domschke.2021}.}. 
Fan et al.~\cite{Fan.2025} combine operational and investment planning for a hydrogen network
and report the reduction in required external storage capacity achieved by exploiting linepack relative to a
steady-state benchmark. While this is the metric closest to an actual storage quantification, it results from cost minimization rather than from a maximization of the storage potential and is only valid for the chosen $\qty{24}{\hour}$ optimization horizon.

To the best of our knowledge, Burlacu et al.~\cite{Burlacu.2019} present the only study whose objective is the
explicit maximization of the linepack storage potential of a gas network and therefore the only study that fits into the third strand.
They quantify the linepack storage potential by an
additional flow rate capacity that can be injected into or withdrawn from selected nodes of a natural gas network. Their analysis, however, is also limited to a $\qty{24}{\hour}$ optimization horizon.

In summary, despite the
growing interest in transient gas network optimization, hardly any approach explicitly quantifies the linepack
storage potential. Where a quantification is provided at all, it is either restricted to a single
pipeline~\cite{Sargent.2025}, a by-product of a cost- or design-driven objective~\cite{Schwele.2019, Fan.2025}, or limited to the conditions of a specific 24-hour period~\cite{Burlacu.2019, Schwele.2019, Fan.2025}.

Moreover, all previous work on transient optimization of gas networks considers short optimization horizons of up to \qty{24}{\hour}~\cite{Zlotnik.2019, OMalley.2021, Shi.2026, Mhanna.2022, Raheli.2025, Schwele.2019, Fan.2025, Burlacu.2019}.
However, linepack storage metrics evaluated on a typical $\qty{24}{\hour}$ horizon only reflect the specific supply and demand situation of the
selected day and cannot be extrapolated to annual storage behavior, since hydrogen supply and demand are both expected to be subject to pronounced seasonal variations~\cite{Lux.2022}.
Additionally, the temporal resolution in existing optimization approaches is typically fixed to hourly~\cite{Mhanna.2022, OMalley.2021, Fan.2025}, quarter-hourly~\cite{Raheli.2025} or $\qty{10}{\min}$~\cite{Burlacu.2019} intervals. However, future hydrogen systems that include hydrogen-fired power plants are expected to exhibit extreme peak loads that may occur only a few times per year~\cite{Lux.2022}. We demonstrate that such peak loads can induce critical transient phenomena that would be overlooked when using an hourly or a quarter-hourly resolution. To avoid optimization results that violate operational constraints, a finer temporal resolution is therefore required—at least during certain critical time periods. Both the long time horizon and the required fine temporal resolution pose significant scalability challenges for the optimization framework that have not been addressed in the existing literature.

\subsection{Our Approach}
As outlined above, the key obstacles to assessing the linepack storage potential are the absence of a
suitable optimization framework that can be solved over a full-year horizon at a sufficiently fine temporal resolution and the lack of metrics that make the potential quantifiable and
comparable to other storage options.

Motivated by this, we transfer metrics established for conventional storage
facilities, like the \emph{flow rate capacity} (injection and withdrawal rate), the \emph{volume capacity}
(working gas volume) and the number of \emph{full-cycle equivalents}, to hydrogen networks. By introducing these metrics for linepack storage and
evaluating them on optimized operating schedules, we can assess the linepack storage potential in terms of
both capacity and operational characteristics, and compare it directly to cavern and underground pipe storage
facilities.

To this end, we formulate an optimization problem for hydrogen networks to maximize the utilization of network-inherent linepack storage capacity by minimizing reliance on external storage. Following established practice, we model the network using the isothermal one-dimensional Euler equations to capture transient linepack changes and enforce operational pressure bounds~\cite{Zlotnik.2019, OMalley.2021, Mhanna.2022, Raheli.2025}. This yields nonlinear PDE-based constraints, which require spatial and temporal discretization. 

To address the computational challenges that arise from solving the nonlinear problem over a full-year optimization horizon combined with a fine temporal resolution, we employ Sequential Convex Programming (SCP). SCP is a generalization of Sequential Linear Programming (SLP), which has 
been shown to offer superior runtime and scalability for gas network optimization compared to general-purpose nonlinear solvers such as IPOPT 
and relaxation-based approaches~\cite{Mhanna.2022, Raheli.2025}. We extend existing SCP-based formulations by introducing an adaptive time-step heuristic that adjusts 
the temporal resolution during the iterative solution process. The key idea is to reduce the problem size by using a coarse resolution where the flow remains quasi-steady, and to refine it only during periods with pronounced transient dynamics. 
Adaptive time-step SCP methods have previously been proposed in the aerospace domain for flight trajectory optimization, e.g.,~\cite{Zhou.2021, Tian.2024, Fu.2026}, where the adaptation of the time mesh is based on the linearization error of the respective iteration. In contrast, we adapt the time step directly according to a discretization error obtained from a high-resolution simulation of the network dynamics. We refer to this approach as \textit{Simulation-Informed Sequential Convex Programming} (SI-SCP) in the following.

To demonstrate that our methodology is well suited to assess the technical linepack storage potential of a hydrogen network, we design a case study based on a network topology from the GasLib standard library~\cite{Schmidt.2017} and energy system scenario data for Germany~\cite{Lux.2022}, which we spatially disaggregate to fit the selected topology. This ensures that the case study realistically reflects operating conditions anticipated for future hydrogen networks, including pronounced winter demand peaks driven by hydrogen-to-power reconversion and highly variable hydrogen production from renewable electricity~\cite{Lux.2022}. We then evaluate the proposed metrics on the resulting operating schedules obtained with SI-SCP.
Furthermore, to examine the relationship between hydrogen transport volumes and linepack storage potential, we perform a sensitivity analysis considering low and high transport volumes in addition to our reference scenario.

In summary, we make the following main contributions to assess the linepack storage potential of hydrogen networks:
\begin{enumerate}
    \item We transfer metrics used to characterize conventional storage facilities (flow rate capacity,
      volume capacity, and full-cycle equivalents) to hydrogen networks, thereby making the linepack storage potential quantifiable and directly comparable to alternative storage options.
    \item We formulate a non-convex optimization problem in which the maximization of linepack storage utilization is the objective itself, subject to PDE-based constraints capturing the 
          transient fluid-mechanical behavior of the hydrogen network.
    \item We solve this problem over a full-year horizon using a novel 
          Simulation-Informed Sequential Convex Programming approach, which 
          drastically reduces the problem size compared to standard SCP while maintaining the same accuracy.
    \item We demonstrate the methodology on the GasLib-40 network topology 
          with hydrogen supply and demand data from energy system analysis 
          and conduct a sensitivity analysis across low- and high-demand 
          scenarios to identify general trends in how linepack storage 
          capacity relates to network load.
\end{enumerate}

The remainder of this paper is organized as follows: We describe the preliminaries for transient hydrogen network modeling in Section~\ref{sec:network_model}, define the metrics to quantify the linepack storage potential in Section~\ref{sec:metrics}, formulate the optimization problem in Section~\ref{sec:optimization}, present the SI-SCP algorithm in Section~\ref{sec:siscp}, describe the case study and evaluate the results in Section~\ref{sec:results} and conclude the paper in Section~\ref{sec:conclusion}.

\subsection*{NOTATION}
Let $\R$, $\R^+$ and $\N$ denote the real, non-negative real and natural numbers, respectively. 
For a finite set $\mc{S}$, $|\mc{S}|$ denotes its cardinality and 
$\mc{S}^{(2)} := \{ \{s_1, s_2\} \subset \mc{S} \mid s_1 \neq s_2 \}$ 
the set of all unordered pairs of its elements. 
Let $I_n$ denote the $n\times n$ identity matrix and 
$0_{m\times n}$ the $m\times n$ zero matrix. Let $\mathbf{1}_n$ and $\mathbf{0}_n$ denote the $n$-dimensional vectors of ones and zeros, respectively.

\section{Transient Hydrogen Network Modeling}\label{sec:network_model}

In this section, we describe the governing partial differential equations (PDEs) of the hydrogen network model. We apply a suitable spatial discretization scheme to derive an ordinary differential equation (ODE) formulation of the network dynamics. 

\subsection{Graph Representation}
We represent the topology of the hydrogen network by a connected and undirected graph $\mc{G}:=(\mc{N},\mc{E})$, with $n \in \N$ nodes and $e \in \N$ edges, where $\mc{N} := \{1, \dots, n\} \subset \N$ denotes the set of nodes and $\mc{E} := \{e_1, \dots, e_e \} \subset \mc{N}^{(2)} $ denotes the set of edges.
The set of nodes is further differentiated into $n_{\mrm{l}}$ load nodes \(\mc{N}_{\mrm{l}} : = \{1, \dots, n_{\mrm{l}}\} \subset \N\)  , with the term load including both supply and demand, and $n_{\mrm{s}} = n - n_{\mrm{l}}$ storage nodes \(\mc{N}_{\mrm{s}} := \{n_{\mrm{l}+1}, \dots, n\} \subset \N\).
The set of edges is further differentiated into $e_{\mrm{p}}$ pipelines \(\mc{E}_{\mrm{p}} := \{e_1, \dots, e_{e_{\mrm{p}}} \} \subset \mc{N}^{(2)}\) and $e_{\mrm{r}} = e - e_{\mrm{p}}$ resistors \(\mc{E}_{\mrm{r}}:= \{e_{e_{\mrm{p}}+1}, \dots, e_{e} \} \subset \mc{N}^{(2)}\). The node-edge-incidence matrix $E \in \R^{|\mathcal{N}| \times |\mathcal{E}|}$ of the network is element-wise defined by
\begin{equation}
    (E)_{ij} = \begin{cases}
        - 1, & \textrm{node} \; i \;  \textrm{is the sink of edge} \, e_j \\
        +1,  & \textrm{node} \; i \;  \textrm{is the source of edge} \, e_j \\
        0, & \textrm{otherwise}.
    \end{cases}
\end{equation}
In the following, $e_{\text{row}}^i$ denotes the row $i \in \mathcal{N}$ of $E$ and $e_{\textrm{col}}^{j}$ denotes the column $j$ for $e_j = (l,r) \in \mathcal{E}$ of $E$.

\subsection{Pipeline Hydraulics}

The hydraulics of gas flow in a pipeline  $e_j  \in \mc{E}_{\mrm{p}}$ are described by the one-dimensional, isothermal, semi-linear Euler equations with constant compressibility, as commonly applied in natural gas network modeling~\cite{Zlotnik.2019, OMalley.2021, Mhanna.2022, Raheli.2025, Malan.2023}. As discussed in~\cite{Herrmann.2024}, the standard simplifying assumptions underlying this formulation are also valid for hydrogen. Therefore, we adopt the same set of equations as the basis for our analysis.
The balance of mass
\begin{equation} \label{eq:euler_1}
        0 =\partial_t p^{j}+\frac{\cs^2}{A^{j}}\partial_{x}
        q^{j},
\end{equation}
the balance of momentum

\begin{equation} \label{eq:euler_2}
0=\partial_tq^{j}+A^{j}\partial_{x}p^{j}+\frac{\lambda^{j}}{2A^{j}D^{j}} \frac{q^{j}|q^{j}|}{\rho^{j}} +  A^{j} g \sin(\theta^{j}) \rho^{j}
\end{equation}
and the equation of state for real gases with constant compressibility
\begin{equation} \label{eq:eq_of_state}
p^{j}=\rho^{j} R_{\mrm{s}} T z = \rho^{j}\cs^2 ,
\end{equation} 
form a system of PDEs, describing the three space and time dependent dynamic states pressure $p^{j}(x,t) \in \R^+$, density $\rho^{j}(x,t) \in \R^+$ and mass flow $ q^{j}(x,t) \in \R $. We consider the isothermal case, hence the temperature $T \in \R^+$ is assumed to be constant in space and time and equal for all pipelines $e_j  \in \mc{E}_{\mrm{p}}$.
The constants $ A^{j}\in \R^+, D^{j}\in \R^+, \; \theta^{j} \in [-\pi/2, \; \pi/2], \; z \in \R^+, \; \cs\in \R^+ \; \textrm{and} \; R_{\mrm{s}}\in \R^+$ denote the pipeline's cross-sectional area, diameter and inclination as well as the fluid's compressibility, isothermal speed of sound and specific gas constant, respectively. We calculate the mass flow dependent friction coefficient $\lambda^{j}(q^{j})\in \R^+$ using the equation of Chen~\cite{Domschke.2021}. The gravitational acceleration is denoted by $g \in \R^+$.

\subsection{Spatial Discretization of the Pipeline Equations}
To solve the system of PDEs~\eqref{eq:euler_1}-\eqref{eq:eq_of_state}, we perform a spatial discretization and obtain a system of ODEs. For this, we use the lumped parameter model described by Malan et al., which uses an analogy from modeling electricity networks to derive a simple spatial discretization with one pressure state $p^{i} \in \R^+$ at each node $i \in \mc{N}$ and one flow state $q^j\in\R$ at each edge $e_j\in\mc{E}$~\cite{Malan.2023}. Let $\mathbf{p} \in \R^{|\mathcal{N}|}$ and $\mathbf{q} \in \R^{|\mathcal{E}|}$ denote the vectors of all node pressures and all edge flows respectively.

Following~\cite{Malan.2023}, spatial discretization of the mass 
balance~\eqref{eq:euler_1} yields
\begin{equation}\label{eq:mass_bal_pimod}
V_{\text{eq}}^i \dot{p}^i =
 q_{\text{ext}}^i - e_{\text{row}}^i \mathbf{q}, \quad
  i \in \mathcal{N}, 
\end{equation}
where $q_{\text{ext}}^i$ denotes the external mass flow entering or 
exiting the network at node~$i$, and
\begin{equation}\label{eq:Veq}
V_{\text{eq}}^i
= e_{\text{row}}^i \, 
   \operatorname{Diag}\!\left[ \frac{L^{j} A^{j}}{2 \cs^2} \right] 
   \left(e_{\text{row}}^i\right)^T,
\quad i \in \mathcal{N}, e_j \in \mathcal{E},
\end{equation}
is the lumped equivalent volume associated with node~$i$. 

Spatial discretization of the momentum 
balance~\eqref{eq:euler_2} gives
\begin{equation}\label{eq:mom_bal_pimod}
    \frac{ L^{j}}{A^{j}} \dot{q}^{j} = 
     - \frac{\lambda^{j}(q^{j}) \, \cs^2 L^{j} \, |q^{j}| q^{j}}{2 D^{j} \left(A^{j}\right)^2 p_\mrm{M}^{j}}
     + \left(e_{\textrm{col}}^{j}\right)^T \mathbf{p}
     - \frac{g L^{j} \sin(\theta^{j})}{\cs^2} p_\mrm{M}^{j}, 
     \quad e_j \in \mathcal{E}_\mrm{p}. 
\end{equation}

The mean pressure $p^{j}_\mrm{M} \in \R^+$ in a pipeline $e_j = (l,r) \in \mathcal{E}_\mrm{p}$ is defined as 
\begin{equation}\label{eq:mean_pressure}
    p^{j}_\mrm{M} = \frac{2}{3} \frac{\left(p^l\right)^3-\left(p^r\right)^3}{\left(p^l\right)^2-\left(p^r\right)^2}.
\end{equation}

\subsection{Non-pipeline Elements and Flow Control}
In our analysis, the flow is not controlled through control valves but through boundary flows $q^i_\mrm{ext}$ at the nodes $ \; i \in \mathcal{N}$. To complete the network description, we define a passive resistor model that serves to replace all non-pipeline elements in the network like compressor stations and control valves. Note that compressor modeling is beyond the scope of this paper.\footnote{We assume that sufficient compression capacity is installed at all feed-in nodes such that feed-in is possible at all times and that intermediate compression is not necessary. This assumption is particularly reasonable during the hydrogen market ramp up, when relatively small amounts of hydrogen are transported in relatively large repurposed natural gas pipelines and hence frictional losses are small.} 
The linear resistor model represents a simplification of the nonlinear resistor model commonly used 
for gas network components~\cite{Koch.2015} and is given as
\begin{equation} \label{eq:valve_equation}
q^{j}=\frac{1}{R}\left(e_{\textrm{col}}^{j}\right)^T \mathbf{p} , \quad e_j \in \mathcal{E}_{\mrm{r}},
\end{equation}
with $R > 0$ being the resistance coefficient. This linearization is analogous to Ohm's law in electrical circuit analysis, with mass flow corresponding to current and pressure to voltage. By expressing resistor flows directly as a linear function of the nodal pressures, they can be substituted into the nodal mass balance, so that the overall system retains a pure ODE structure without additional algebraic constraints.

\subsection{ODE Formulation of the Network Dynamics}

The algebraic equations~\eqref{eq:mean_pressure} 
and~\eqref{eq:valve_equation} can be substituted directly into the 
differential equations~\eqref{eq:mom_bal_pimod} 
and~\eqref{eq:mass_bal_pimod}, respectively. Consequently, the 
fluid-mechanical network 
model~\eqref{eq:mass_bal_pimod}--\eqref{eq:valve_equation} reduces to a 
nonlinear system of ODEs of the general form
\begin{equation}\label{eq:network_ode}
    \dot{x} = f(x, u),
\end{equation}
with the state vector 
$x = [\mathbf{p}^T,\, \mathbf{q}^T]^T 
\in \R^{|\mathcal{N}|+|\mathcal{E}|}$ 
and the control input 
$u = \mathbf{q}_\mrm{ext} \in \R^{|\mathcal{N}|}$ collecting all 
external mass flows. The input $u(t)$ is parameterized as a zero-order 
hold signal with hourly changing values.

The ODE system~\eqref{eq:network_ode} serves as a dynamic constraint in 
the optimization problem formulated in 
Section~\ref{sec:optimization}. Furthermore, when all external mass 
flows $\mathbf{q}_\mrm{ext}(t)$ are known, 
\eqref{eq:network_ode} reduces to the initial value problem stated in 
Problem~\ref{prob:sim_prob} below. In our proposed framework, we solve Problem~\ref{prob:sim_prob} to assess the accuracy of the 
optimization results and to determine the step sizes in our adaptive 
time-stepping scheme (see Section~\ref{sec:siscp}).

\begin{problem}[Initial Value Problem]\label{prob:sim_prob}
Given the network equations~\eqref{eq:network_ode}, the external mass 
flows $u(t) = \mathbf{q}_\mrm{ext}(t)$ for all 
$t \in [0,\, T]$, with $T \in \R^+$ denoting the simulation (and later 
optimization) horizon, solve the initial value problem
\begin{subequations}\label{eq:ivp}
\begin{align}
    \dot{x}(t) &= f\!\left(x(t),\, u(t)\right),
        \quad t \in [0,\, T], \label{eq:ivp_ode} \\
    x(0) &= \begin{bmatrix}
        \mathbf{p}^* \\ \mathbf{q}^*
    \end{bmatrix}. \label{eq:ivp_ic}
\end{align}
\end{subequations}
\end{problem}
The initial condition $x(0) = x^* = [(\mathbf{p}^*)^T,\, 
(\mathbf{q}^*)^T]^T$ is obtained as a steady state by solving
\begin{subequations}\label{eq:steady_state}
\begin{align}
    \mathbf{0}_{|\mathcal{N}|+|\mathcal{E}_{\mrm{p}}|} 
    &= f(x^*, u^*), \\
    x^{*}_i &= p_\mrm{ext}, \\
    u^{*}_j &= q^j_\mrm{ext}(0), \quad j\neq i, j\in \mc{N},
\end{align}
\end{subequations}
where the externally given pressure $p_\mrm{ext} \in \R^+$ at a selected node $i \in \mc{N}$ fixes the otherwise undetermined pressure level of the network, while the corresponding injection $u^*_i$ follows from the overall mass balance.

To ensure both accuracy and computational efficiency, we solve 
Problem~\ref{prob:sim_prob} using implicit ODE solvers from the 
\texttt{DifferentialEquations.jl} library~\cite{Rackauckas.2017} in 
Julia. These solvers employ high-order discretization schemes with 
adaptive time stepping and efficiently exploit the sparsity structure of 
the system. The model is implemented in the 
\texttt{NetworkDynamics.jl}~\cite{Lindner.2021} framework, which allows 
us to construct the right-hand side of~\eqref{eq:network_ode} in a 
modular fashion: the dynamics of pipelines, resistors, load nodes, and 
storage nodes are defined individually and interconnected according to 
the network topology.

\section{Linepack Storage Metrics} \label{sec:metrics}
As outlined in Section~\ref{sec:literature}, the existing literature provides no metric that quantifies the linepack storage potential of a gas network in a way that is comparable to dedicated storage facilities. 
Conventional hydrogen storage facilities are typically characterized by injection and withdrawal rates, working gas volume and \textit{full-cycle equivalents} (e.g.,~\cite{LudwigBolkowSystemtechnikGmbH.2022}). In the sequel, we generalize the injection and withdrawal rates as maximum and minimum \textit{flow rate capacities} and the working gas volume as a \textit{volume capacity}, and subsequently transfer these metrics to the linepack storage of hydrogen networks.

Pipeline linepack is the mass of gas that is contained within one pipeline or resistor $e_j \in \mathcal{E}$ at a time $t$~\cite{Raheli.2025}:
\begin{equation} \label{eq:linepack_pipe}
m^{j}(t)=A^{j}\int_0^{L^{j}} \rho^{j}(x,t)\, dx=\frac{A^{j} }{R_\mrm{s} T  z} \int_0^{L^{j}} p^{j}(x,t) \, dx.
\end{equation} 
Consequently, the linepack of the network at a time $t$ is the total mass of gas in the network, i.e., 
\begin{equation} \label{eq:linepack_net}
m_\mrm{tot}(t) = \sum_{e_j \in \mathcal{E}} m^{j}(t).
\end{equation} 
The linepack~\eqref{eq:linepack_net} alone does not quantify an operationally useful storage capacity of the network. The network can only serve as storage if the linepack can be varied without violating operational constraints. We therefore define the \textit{linepack storage flow rate capacity} as the 
artificial network-wide flow rate $q_\mrm{linepack}(t):=\frac{dm_\mrm{tot}}{dt}(t)$. 
From the mass balance~\eqref{eq:mass_bal_pimod} for the entire network it follows that
\begin{equation} \label{eq:linepack_storage}
q_\mrm{linepack}(t) = \frac{dm_\mrm{tot}}{dt}(t) = \sum_{i \in \mathcal{N}} q^{i}_\mrm{ext}(t)\begin{cases}
    <0, & \text{network discharge} , \\
    >0, & \text{network charge}, \\
    =0 .
\end{cases}
\end{equation} 
The relation~\eqref{eq:linepack_storage} directly reveals why steady-state hydrogen network models are
unsuited for analyzing the linepack storage flow
rate capacity: there, $\sum_{i \in \mathcal{N}} q^{i}_\mrm{ext}(t) = 0$ holds by definition, enforced by at least one node $i \in \mathcal{N}$ with a pressure boundary condition $p^i = p_\mrm{ext}$. 
In a transient model, by contrast, all boundary conditions can be mass flow boundary conditions as described in \eqref{eq:mass_bal_pimod}, since the pressure can always be inferred from the initial condition in the transient model. 

We can split the mass flow boundary conditions in~\eqref{eq:mass_bal_pimod} according to the type of node:
\begin{equation}\label{eq:q_ext}
    q^i_\mrm{ext}(t) = \begin{cases}
    q^{i}_\mrm{load}(t), & i \in \mathcal{N}_\mrm{l} , \\
    -q^{i}_\mrm{stor}(t), & i \in \mathcal{N}_\mrm{s} . \\
\end{cases}
\end{equation}
The load flows $q^{i}_\mrm{load}(t)$ are determined by electricity and 
hydrogen market clearings and thus considered unalterable in our 
analysis. Their sum 
$\sum_{i \in \mathcal{N}_\mrm{l}} q^{i}_\mrm{load}(t) =: -q_\mrm{res}(t) \in \R$ 
defines the residual demand, which, from~\eqref{eq:linepack_storage}, 
satisfies
\begin{equation}\label{eq:qres}
    q_\mrm{res}(t) = -q_\mrm{linepack}(t) - \sum_{s \in \mathcal{N}_\mrm{s}} q^{s}_\mrm{stor}(t).
\end{equation}

Since $q_\mrm{res}(t)$ is fixed, equation~\eqref{eq:qres} reveals 
that the distribution of the residual between linepack storage 
$q_\mrm{linepack}(t)$ and external storage flows 
$q^{s}_\mrm{stor}(t)$ constitutes a degree of freedom that can be 
exploited within the operational constraints of the network. 
To determine $q_\mrm{linepack}(t)$ from~\eqref{eq:qres}, an additional condition for $\sum_{s \in \mathcal{N}_\mrm{s}} q^{s}_\mrm{stor}(t)$ is required. For obtaining such a condition, we introduce two closure rules that represent the extreme cases of the admissible range of linepack storage utilization: a Benchmark Scenario, in which the entire 
residual is covered by external storage (i.e.\ zero linepack 
utilization), and an Optimized Scenario, in which linepack 
utilization is maximized by minimizing external storage usage.

\textbf{Benchmark Scenario.}
In this scenario, we assume that the external mass flows are balanced at 
every point in time, i.e.\ 
$\sum_{i \in \mathcal{N}} q^{i}_\mrm{ext}(t) = 0, \; \forall t$. 
Such a condition arises naturally when scenario data is derived from 
energy system optimizations that do not account for linepack storage. It follows from~\eqref{eq:q_ext} that
\begin{equation}\label{eq:min_linepack_adapted}
    \sum_{s \in \mathcal{N}_\mrm{s}} q^{s}_\mrm{stor}(t) = \sum_{l \in \mathcal{N}_\mrm{l}} q^{l}_\mrm{load}(t) = -q_\mrm{res}(t),\; \forall t.
\end{equation}
Combining this with~\eqref{eq:qres} yields $q_\mrm{linepack}(t)=0, \; \forall t$.

\textbf{Optimized Scenario.}
In this scenario, the external-storage flows are determined 
endogenously by solving
\begin{equation}\label{eq:max_linepack_adapted}
    \min_{q^{s}_\mrm{stor}} \;
    \int_{t_0}^T \sum_{s \in \mathcal{N}_\mrm{s}}
    \left(q^{s}_\mrm{stor}(t)\right)^2 dt,
\end{equation}
subject to the operational constraints of the hydrogen network. Because the residual demand is fixed, minimizing the external-storage 
flows shifts as much of the balancing task as operationally feasible 
to linepack storage.
The separable, strictly convex quadratic cost function discourages large 
storage flows irrespective of their sign and prevents large opposing 
flows at different nodes from canceling in the objective. The full 
optimization problem incorporating~\eqref{eq:max_linepack_adapted} is 
described in Section~\ref{sec:optimization}.

Once we get $q_\mrm{linepack}(t)$ from~\eqref{eq:qres} and~\eqref{eq:min_linepack_adapted} or~\eqref{eq:max_linepack_adapted}, respectively, we can calculate an equivalent working gas volume at standard conditions of the network $V_\mrm{linepack}(t)$ by integrating
\begin{equation}\label{eq:storage_eq}
    \frac{dV_\mrm{linepack}(t)}{dt} = \frac{q_\mrm{linepack}(t)}{\tilde{\rho}},
\end{equation}
with $\tilde{\rho}$ denoting the standard density of hydrogen.
Based on the working gas volume trajectory, we define the \textit{linepack storage volume capacity} over an annual horizon $[0, T]$ with $T = \qty{8760}{\hour}$ as
\begin{equation}\label{eq:vol_capa}
     \Delta V_\mrm{linepack} = \max_{t\in[0, T]}V_\mrm{linepack}(t) - \min_{t\in[0, T]} V_\mrm{linepack}(t).
\end{equation}
Note that $\Delta V^\mrm{Benchmark}_\mrm{linepack} = 0$ for the Benchmark scenario. Analogous to \eqref{eq:storage_eq} and \eqref{eq:vol_capa} we can also calculate the working gas volume $V_\mrm{stor}^{s}(t)$ and the volume capacity $\Delta V_\mrm{stor}^{s}$ of a cavern storage facility $s \in \mathcal{N}_\mrm{s}$.

Another important metric to characterize a storage unit in terms of short- or long-term operation is the number of \textit{full-cycle equivalents}, a concept commonly used in storage analysis, e.g. in~\cite{LudwigBolkowSystemtechnikGmbH.2022}. It can be defined analogously for linepack and cavern storage as the annual throughput divided by the volume capacity, i.e.,
\begin{equation}\label{eq:fce}
    \mrm{FCE} = \frac{1}{2 \Delta V_{mode} \;\tilde{\rho}}\int_{0}^{T} |q_{mode}(t)|dt, \quad mode \in \{\mrm{linepack}, \mrm{stor}\}.
\end{equation}
Since the metrics~\eqref{eq:vol_capa} and~\eqref{eq:fce} inherently require a full-year evaluation horizon, 
this further motivates the annual scope of the optimization problem 
introduced below.

\section{Optimization Problem} \label{sec:optimization}
We derive a nonlinear optimization problem incorporating the dynamic network model~\eqref{eq:network_ode} as a collocation constraint. This requires temporal discretization of~\eqref{eq:network_ode}, additional operational constraints, and an appropriate objective function.

\subsection{Temporal Discretization}
Since standard gradient-based solvers cannot handle differential-equation constraints, we apply a temporal discretization to~\eqref{eq:network_ode} to obtain a fully discretized collocation constraint.
In the following, we distinguish between the temporal discretization of the system states $x$ and the control inputs $u$. The control inputs are discretized hourly, since load-flow data is typically available at hourly resolution (e.g.\ from market clearings or energy system models). As linear interpolation between hourly values would smooth out variations and underestimate the dynamic operation, we employ a zero-order hold representation of $u$, i.e., constant control inputs within each hourly interval.
In contrast, the temporal discretization of the states $x$ may be finer than hourly. In fact, it is shown in the sequel that an hourly resolution cannot accurately capture the system dynamics in all cases.

Let the optimization horizon $[t_0, T]$ be discretized by a possibly non-uniform sequence of discrete time steps $t_k$ at which the system states $x_k = x(t_k)$ are evaluated. We define $t_k$ by
\begin{equation}
  t_{k+1} = t_k + \Delta t_k, \qquad \Delta t_k > 0, \quad k = 0,\dots,K-1, \, K \in \N,
\end{equation}
with $t_0$ denoting the initial time and $t_K = T$ the end time of the horizon.
Additionally, with $H\in\N$,  we define an hourly subset $\{t_h\}_{h=0}^H \subset \{t_k\}_{k=0}^K$ on which the control inputs $u_h = u(t_h)$ are evaluated with
\begin{equation}
  t_h= t_0 + h \cdot 1\,\text{h}, \qquad h = 0, \dots, H
\end{equation}
and $t_H \le t_K$.
We choose $k$ and $h$, so that there exists a strictly increasing mapping
\begin{equation}
  k : \{0,\dots,H\} \to \{0,\dots,K\},
  \quad\text{such that}\quad
  t_{k(h)} = t_h \quad \forall\, h.
\end{equation}

With this we define the piecewise constant control input $u$ as follows
\begin{equation}\label{eq:pw_constant_control}
  u_k = u_h \quad \forall  t_k \in [t_h,\, t_{h+1}).
\end{equation}

\subsection{Constraints}
\paragraph{Collocation constraints} With the discrete time steps 
defined above, we follow~\cite{Burlacu.2019} and apply an implicit Euler scheme for temporal discretization. Discretizing~\eqref{eq:network_ode} yields
\begin{equation}\label{eq:network_ode_discretized}
    \frac{x_{k+1} - x_k}{\Delta t_{k}}  =  f(x_{k+1}, u_{k+1}), \quad k=0,..,K-1,
\end{equation}
which describes a set of nonlinear collocation constraints.

\paragraph{Operational constraints} Besides the collocation constraints~\eqref{eq:network_ode_discretized}, 
we define two additional sets of constraints.
First, all network pressures must remain within the operational limits $p_\mrm{min}\in \R^+$ and $p_\mrm{max}\in \R^+$. Let 
$\underline{p} = p_\mrm{min}\,\mathbf{1}_{|\mathcal{N}|}$ and 
$\overline{p} = p_\mrm{max}\,\mathbf{1}_{|\mathcal{N}|}$ denote the 
corresponding bound vectors. We enforce
\begin{equation}\label{eq:pressure_box}
  \underline p \;\le\;  C_\mrm{p}\,x_k \;\le\; \overline p,
  \qquad k = 0,\dots,K,
\end{equation}
where
$C_\mrm{p} := \begin{bmatrix} I_{|\mathcal{N}|} & 
0_{|\mathcal{N}|\times |\mathcal{E}_\mrm{p}|} \end{bmatrix}$
is the pressure selection matrix.

Second, we fix the external mass flows $q^{i}_\mrm{load}$ at the load nodes $i \in \mc{N}_{\mrm{l}}$ to their 
prescribed values. As discussed in Section~\ref{sec:metrics}, 
these flows represent the residual demand~\eqref{eq:qres} that is determined 
externally, e.g.\ by market clearings or contractual 
obligations, and cannot be influenced by the network or storage 
operator. They therefore enter the optimization as a fixed 
disturbance. 
Since the load nodes occupy the first $n_\mrm{l}$ indices by 
construction, the selection matrix 
$C_{\mathcal{N}_\mrm{l}} = 
\begin{bmatrix} I_{|\mathcal{N}_\mrm{l}|} & 0_{|\mathcal{N}_\mrm{l}| \times |\mathcal{N}_\mrm{s}|} 
\end{bmatrix}$ 
extracts the load-node entries of~$u_h$. Given the vector of prescribed load 
flows 
$\mathbf{q}_{\mrm{load},h} \in \mathbb{R}^{|\mathcal{N}_\mrm{l}|}$, we impose
\begin{equation}\label{eq:load_match}
    C_{\mathcal{N}_\mrm{l}}\,u_h
    \;=\;
    \mathbf{q}_{\mrm{load},h},
    \qquad h = 0,\dots,H.
\end{equation}
Consequently, the remaining degrees of freedom of the optimizer 
reduce to $|\mathcal{N}_\mrm{s}|$~storage flows 
$q^s_\mrm{stor}(t_h)$, $s \in \mathcal{N}_\mrm{s}$.

\subsection{Objective Function}
As motivated in Section~\ref{sec:metrics}, minimizing the squared 
external storage flows maximizes linepack utilization while 
penalizing large opposing flows at different storage 
nodes~\eqref{eq:max_linepack_adapted}. Discretizing this objective 
under the piecewise-constant control assumption~\eqref{eq:pw_constant_control} yields
\begin{equation}\label{eq:cost-function}
    \ell(u_h) 
    = \sum_{h=1}^{H} \left\| C_{\mathcal{N}_\mrm{s}}\, u_h \right\|_2^2
    = \sum_{h=1}^{H} \sum_{s \in \mathcal{N}_\mrm{s}} 
    \left(q^s_{\mrm{stor},h}\right)^2,
\end{equation}
with $C_{\mathcal{N}_\mrm{s}} = \begin{bmatrix} 0_{|\mathcal{N}_\mrm{s}| \times |\mathcal{N}_\mrm{l}|} 
& I_{|\mathcal{N}_\mrm{s}|} \end{bmatrix}$ extracting the storage-node entries of~$u_h$.

Depending on the pressure level used for the initial steady-state, the linepack storage will come with an initial ``state of charge''. To prevent the optimizer from exploiting this by ``emptying'' the network at the end of the optimization horizon, we introduce a terminal cost term $\ell_\mrm{tc}(x_{k=K})$: 
\begin{equation}\label{eq:terminal-cost}
    \ell_\mrm{tc}(x_K) 
    = \alpha \left\| C_{\mathcal{N}_\mrm{s}}\, C_\mrm{p}\, x_K 
      - C_{\mathcal{N}_\mrm{s}}\, C_\mrm{p}\, x_0 \right\|_2^2
    = \alpha \sum_{s \in \mathcal{N}_\mrm{s}} 
      \left(p^s_K - p^s_0\right)^2,
\end{equation}
with $\alpha \in \R^+$ being a scalar weight and $p_0 = C_\mrm{p}x_0 \in \R^{|\mathcal{N}|}$ being the constant initial state.
At the final time step, this term keeps the pressures at a selected subset of critical nodes close to their initial steady-state values, without requiring the full network to return to steady state. In our case study, we choose the storage nodes $\mathcal{N}_\mrm{s}$ as the critical subset, but other network-specific choices are equally feasible.

With the constraints and cost functions defined, we can formulate the following optimization problem.

\begin{problem}[Nonlinear Optimization Problem]\label{prob:opt_prob_nl}
\begin{subequations}\label{eq:opt_prob_nl}
At time $k=h=0$, initial states $x_{0}$ and inputs $u_{0}$ are given. Now, consider~\eqref{eq:network_ode_discretized}-\eqref{eq:terminal-cost} and solve
\begin{alignat}{2}
    \min_{u_h} \quad
    & \left(\ell(u_h)
      \;+\; \ell_\mrm{tc}(x_K)\right)
    \label{eq:objective} \\
    \textrm{s.t.} \quad
    & x_{k+1} = x_{k} + \Delta t_k f(x_{k+1},\, u_{k+1}),
    &\quad & k = 0,\dots,K\!-\!1,
    \label{eq:opt_dyn} \\
    & \underline p \;\le\;  C_\mrm{p}\,x_k \;\le\; \overline p,
    &\quad & k = 1,\dots,K,
    \label{eq:opt_pbox} \\
    & C_{\mathcal{N}_\mrm{l}}\,u_h
      = \mathbf{q}_{\mrm{load},h},
    &\quad & h = 0,\dots,H.
    \label{eq:opt_load}
\end{alignat}
\end{subequations}
\end{problem}

\section{Simulation-Informed Sequential Convex Programming (SI-SCP)}\label{sec:siscp}

Problem \ref{prob:opt_prob_nl} harbors nonlinearity in two components: the nonlinear collocation constraint \eqref{eq:opt_dyn} and the quadratic objective function \eqref{eq:objective}. Since the quadratic objective is convex, linearizing the collocation constraint yields a convex optimization problem (Section~\ref{sec:scp_lin}). The idea of standard SCP is to solve a series of convex problems, each linearized around the solution of the previous iteration, until convergence to the solution of the nonconvex problem~\cite{OMalley.2021, Mhanna.2022, Raheli.2025}. We combine this idea with a simulation-informed adaption of the temporal resolution within each iteration (Section~\ref{sec:adapt_dt}) and describe the termination criteria, warm-start options and formal algorithm for the SI-SCP approach (Section~\ref{sec:siscp_alg}).

\subsection{Linearization of the Network Equations}\label{sec:scp_lin}
First-order Taylor approximation is standard in SLP approaches for 
gas networks~\cite{OMalley.2021, Mhanna.2022, Raheli.2025}. Typically, the linearization is performed 
with respect to the state~$x$ only and can be applied 
either solely to the nonlinear friction term in the momentum 
equation~\eqref{eq:mom_bal_pimod}~\cite{OMalley.2021, Raheli.2025} or to the full ODE 
system via its Jacobian~\cite{Mhanna.2022}. In our 
approach, we linearize the full ODE system~\eqref{eq:network_ode} with respect to both~$x$ 
and~$u$, similar to an approach used in flight trajectory planning~\cite{Tian.2024}, and obtain the required Jacobians by forward-mode automatic 
differentiation using the \texttt{ForwardDiff.jl} library~\cite{Revels.2016}.
With reference trajectories $\hat{x}_{k} \in \R^{(|\mathcal{N}|+|\mathcal{E}_{\mrm{p}}|)} $ and $\hat{u}_{k} \in \R^{|\mathcal{N}|}$ for $ k \in (0,...,K) $ given either as an initial guess or as the solution of the previous SCP iteration, the Jacobians of $f$ are given as
\begin{equation}\label{eq:Jacobian}
\begin{split}
A _k:= & \frac{\partial f}{\partial x}\bigg|_{x=\hat{x}_{k},\,u=\hat{u}_{k}}, \\
B_k := & \frac{\partial f}{\partial u}\bigg|_{x=\hat{x}_{k},\,u=\hat{u}_{k}}.
\end{split}
\end{equation}
With this, the linearization of the fully discretized collocation constraint in~\eqref{eq:network_ode_discretized} reads
\begin{equation}\label{eq:network_ode_discretized_Taylor}
   \frac{x_{k+1} - x_k}{\Delta t_{k}} = 
f\big|_{x=\hat{x}_{k+1},\,u=\hat{u}_{k+1}}
+ A_{k+1}
  \bigl(x_{k+1} - \hat{x}_{k+1}\bigr)
+ B_{k+1}
  \bigl(u_{k+1} - \hat{u}_{k+1}\bigr)
, \quad k=0,..,K-1.
\end{equation}
The Taylor-series approximation~\eqref{eq:network_ode_discretized_Taylor} 
is performed around a full reference trajectory rather than a single 
operating point, yielding a distinct linear model per time step. The so-derived Convex Optimization Problem~\ref{prob:opt_prob} is solved in a series of iterations $a \in 1,\dots,a_{\mrm{max}}$, where $a_{\max} \in \mathbb{N}$ denotes the maximum number of iterations.
\begin{problem}[Convex Optimization Problem at Iteration a]\label{prob:opt_prob}
\begin{subequations}\label{eq:opt_prob}
At time $k=h=0$, initial states $x^{a}_{0}$ and inputs $u^{a}_{0}$ are given. Now, consider~\eqref{eq:pressure_box}-\eqref{eq:terminal-cost},~\eqref{eq:network_ode_discretized_Taylor} and solve
\begin{alignat}{2}
    \min_{u^{a}_h} \quad
    & \left(\ell(u^{a}_h)
      \;+\; \ell_\mrm{tc}(x^{a}_K)\right)
    \label{eq:objective_convex} \\
    \textrm{s.t.} \quad
    & x^{a}_{k+1} = x^{a}_{k} + \Delta t^{a}_k \; f\big|_{x=\hat{x}^{a}_{k+1},\,u=\hat{u}^{a}_{k+1}} &\quad & \nonumber\\
    & + \Delta t^{a}_k \; A^{a}_{k+1}
    \bigl(x^{a}_{k+1} - \hat{x}^{a}_{k+1}\bigr) & \quad & \nonumber\\
    & + \Delta t^{a}_k \; B^{a}_{k+1}
    \bigl(u^{a}_{k+1} - \hat{u}^{a}_{k+1}\bigr)
    &\quad & k = 0,\dots,K^{a}\!-\!1,
    \label{eq:opt_dyn_convex} \\
    & \underline p \;\le\;  C_\mrm{p}\,x^{a}_k \;\le\; \overline p,
    &\quad & k = 1,\dots,K^{a},
    \label{eq:opt_pbox_convex} \\
    & C_{\mathcal{N}_\mrm{l}}\,u^{a}_h
      = \mathbf{q}_{\mrm{load},h},
    &\quad & h = 0,\dots,H.
    \label{eq:opt_load_convex}
\end{alignat}
\end{subequations}
\end{problem}

\subsection{Simulation-Informed Time Step Adaption}\label{sec:adapt_dt}
\begin{figure}[h]
\centering
\includegraphics[width=1\textwidth]{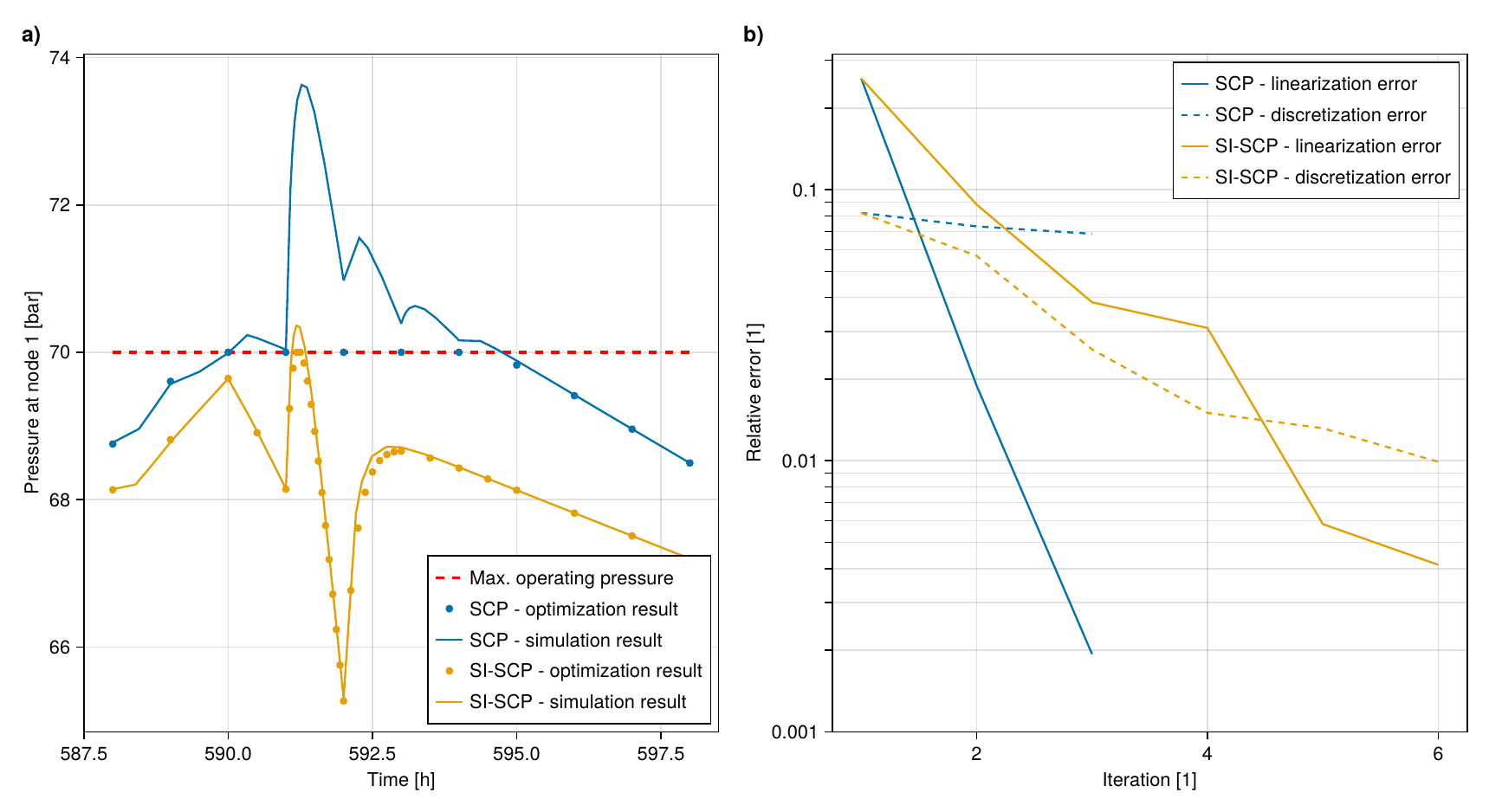}
\caption{a) Pressure transient calculated by SCP and SI-SCP algorithms and validated with simulation results. b) Relative linearization and discretization errors for each iteration of the SCP and SI-SCP algorithms.}
\label{fig:lin-and-disc-error}
\end{figure}

What distinguishes the SI-SCP algorithm from standard SCP algorithms in the gas network literature is that the temporal resolution $\Delta t^{a}_k$ and consequently the number of time steps $K^{a}$ in Problem~\ref{prob:opt_prob} are not fixed but adapted within each iteration $a$. We illustrate why this is necessary with a small three-node-network example described in~\ref{sec:appendix_A}. We solve the Optimization Problem~\ref{prob:opt_prob_nl} for the example scenario using a standard SCP algorithm described in~\ref{sec:appendix_B} with a fixed time step $\Delta t^{a}_k = \qty{1}{\hour}, \, \forall a, \forall k $. The results of the SCP optimization are shown in Figure~\ref{fig:lin-and-disc-error} a). It can be seen that the temporal resolution of the hydrogen network dynamics can have a strong impact on the quality of the optimization result. For the considered scenario, we observe a maximum error of \qty{3.6}{\bar} compared to the solution of the Simulation Problem~\ref{prob:sim_prob}. This relatively large deviation results from a pronounced transient that is largely missed due to the hourly temporal resolution of the optimization. Moreover, this error occurs at a time when the network is operated at the maximum operating pressure (see Figure~\ref{fig:lin-and-disc-error} a)). It is therefore particularly critical, as it leads to a violation of the operating constraints given in \eqref{eq:pressure_box}.

Standard SCP focuses on the linearization error, which can be measured for each iteration $a$, e.g., using the infinity norm
\begin{equation}
   e^{a}_\mrm{lin} = || x^a_k - \hat{x}^a_k||_{\infty}, \quad k \in 1,...,K^a.
\end{equation}
If the actual state is equal to the linearization state at every time step, i.e., 
$x^a_k = \hat{x}^a_k$, the linearized model is accurate and the linearization error vanishes.
As shown in Figure~\ref{fig:lin-and-disc-error} b), the linearization error of the SCP algorithm decreases substantially within three iterations, indicating that the solution of the Convex Optimization Problem~\ref{prob:opt_prob} converges to that of the Nonlinear Optimization Problem~\ref{prob:opt_prob_nl}. However, this holds only for the solution at the optimization time steps $k\in 1,...,K^{a}$. As shown above, solving the Simulation Problem~\ref{prob:sim_prob} at a much finer temporal resolution reveals that the system behavior can deviate strongly in between these time steps. Let $x^{\mrm{sim},a}_l$ denote the simulation result obtained by using $u^{\mrm{opt},a}_h$ as boundary flows, evaluated at the simulation time steps $l \in 1,...,L$, where $t_L = t_{K^{a}} = T$ but $L\gg K^{a}$. Furthermore, let $x^{\mrm{opt},a}_l$ denote the optimization result in iteration $a$ interpolated to the simulation time steps $l$. With this, we define a second error 
\begin{equation}\label{eq:disc_error}
    e^{a}_{\mrm{disc}} = || x^{\mrm{opt},a}_l - x^{\mrm{sim},a}_l||_{\infty}, \quad l \in 1,...,L,
\end{equation}
for each iteration $a$, which we primarily interpret as a discretization error\footnote{It should be noted, however, that $x^{\mrm{opt},a}_l$ refers to the linearized model whereas $ x^{\mrm{sim},a}_l$ refers to the nonlinear model. Therefore the discretization error may still be superimposed by linearization errors, especially in the early iterations.}. Since Problems~\ref{prob:opt_prob_nl} and~\ref{prob:sim_prob} use different temporal resolutions, their solutions cannot coincide, and $e^{a}_{\mrm{disc}}$
does not converge (see Figure~\ref{fig:lin-and-disc-error} b))).

To reduce the discretization error and satisfy the pressure constraints~\eqref{eq:pressure_box} within a given tolerance, a uniformly fine discretization across the entire horizon would drastically increase the problem size and computational burden. Instead, we propose an adaptive strategy that uses the simulation result to identify the intervals where the discretization error exceeds the tolerance, and refines the temporal resolution only there. We refer to this approach as \textit{Simulation-Informed Sequential Convex Programming (SI-SCP)}.

The focus on the discretization error constitutes the main difference between our approach and an existing mesh-adaptive SCP method developed for aerospace applications. The works in~\cite{Zhou.2021, Fu.2026} likewise compare the solution of the Convex Optimization Problem~\ref{prob:opt_prob} to a (local) integration of the nonlinear ODE Problem~\ref{prob:sim_prob} to identify regions in which the time mesh must be refined. However, they evaluate this error only at the optimization time steps $k$, which effectively amounts to assessing the linearization error~\cite{Zhou.2021, Fu.2026}. In contrast, we evaluate the discretization error directly by exploiting the fine temporal resolution of the employed ODE solver\footnote{Further differences are that we consider a global rather than a local error, and that our algorithm does not remove mesh points when the calculated error is low.}.

More specifically, to simplify the approach and improve convergence, we evaluate the discretization error only with respect to the pressure states, since these are critical for the operational constraints. We then identify the hours $h$ in which the discretization error exceeds a predefined absolute pressure tolerance $\mrm{tol}_{\mrm{p}}$, i.e., $e^{a}_{\mrm{disc}, h} > \mrm{tol}_{\mrm{p}}$ with
\begin{equation}
    e^{a}_{\mrm{disc}, h} = ||C_{\mrm{p}} x^{\mathrm{opt},a}_{l} 
        - C_{\mrm{p}}  x^{\mathrm{sim},a}_{l} \|_{\infty}, \,\, \text{with} \, \, t_l \in [t_h,t_{h+1}).
\end{equation}
Whenever the pressure error exceeds the tolerance $\mathrm{tol}_{\mrm{p}}$ in a given hour, the resolution during that hour is doubled~\eqref{eq:double_temp_res}: from hourly to half-hourly in the first iteration, and further in later iterations if the hour $h$ remains problematic, i.e.,
\begin{equation}\label{eq:double_temp_res}
    \Delta t^{a+1}_{k} = 0.5\,\Delta t^{a}_{k} \quad \forall k \, \,\text{with} \,\, t_k \in [t_h,t_{h+1})
\end{equation}
As shown in Figure~\ref{fig:lin-and-disc-error} a), the finer temporal resolution during critical transient phenomena enables the SI-SCP algorithm to satisfy the operational constraints within a small given tolerance (here $\mrm{tol}_{\mrm{p}}=\qty{0.7}{\bar}$). At the same time, a coarse hourly resolution can be retained once the transient phenomena have subsided (see $t \in [\qty{595}{\hour}, \qty{598}{\hour}]$ in Figure~\ref{fig:lin-and-disc-error} a)). Figure~\ref{fig:lin-and-disc-error} b) shows that both the linearization and the discretization error decrease as a result of the adaptive time stepping. This comes at the cost of an increased number of iterations and the additional computational burden of solving Problem \ref{prob:sim_prob} in each iteration. Having illustrated the benefits and trade-offs of the adaptive time stepping, we now formalize the overall SI-SCP procedure in the next section.

\subsection{SI-SCP Algorithm}\label{sec:siscp_alg}

Problem~\ref{prob:opt_prob} depends on the reference trajectories 
$\hat{x}_k$ and $\hat{u}_h$ through the linearized dynamics 
constraint~\eqref{eq:opt_dyn_convex}. In the SI-SCP framework, we 
re-solve this convex problem iteratively for 
$a \in \{1,\dots,a_{\mrm{max}}\}$. For~$a \geq 2$, the 
reference trajectories are set to the preceding optimal solution, $\hat{x}_k^{a} = x_k^{\mrm{sim}, a-1}$ and 
$\hat{u}_h^{a} = u_h^{a-1}$, and Problem~\ref{prob:opt_prob} is 
re-solved with the updated linearization. For the first iteration 
($a=1$), no prior solution is available; the construction of 
suitable initial reference trajectories $\hat{x}_k^1$ and 
$\hat{u}_h^1$ is described in the \emph{Warmstart} paragraph below. 
The iterative process continues until one of the following 
termination criteria is satisfied.

\paragraph{Termination Criteria}
In the SI-SCP approach, the iteration terminates either when the maximum number of iterations $a_{\mrm{max}}$ is reached, i.e.,
\begin{equation}
    a = a_{\mrm{max}},
\end{equation}
or once the optimization result has sufficiently converged. We assess convergence using both the linearization and the discretization error. For the linearization error, we consider the input $u$, since it represents the storage flow, which is directly relevant to the linepack storage capacity.
The sequential convexification terminates only if both tolerances are met, $\mathrm{tol}_{\mrm{u}}$ for the linearization error and $\mathrm{tol}_{\mrm{p}}$ for the discretization error, i.e., 
\begin{equation}
    || u^{\mrm{opt}}_h - \hat{u}_h||_{\infty} < \mathrm{tol}_{\mrm{u}} \quad \wedge \quad || C_{\mrm{p}} x^{\mrm{opt}}_k - C_{\mrm{p}} x^{\mrm{sim}}_k||_{\infty} < \mathrm{tol}_{\mrm{p}}.
\end{equation}

\paragraph{Warmstart}
Prior work on SCP for gas networks emphasizes that a high-quality initial guess $\hat{x}_{k}^1$ can substantially improve algorithmic performance~\cite{Mhanna.2022, Raheli.2025}. In our setting, a natural choice exists for both the initial reference control $\hat{u}_{h}^1$ and the corresponding reference states $\hat{x}_{k}^1$. We construct $\hat{u}_{h}^1$ from the balanced storage-flow profile of the benchmark scenario, which is derived from \eqref{eq:qres} with $q_{\mrm{linepack},h}=0 \;\forall h $. The storage-related components of  $\hat{u}^1_h$ are therefore given by
\begin{equation}\label{eq:warmstart}
    C_{\mathcal{N}_\mrm{s}}\hat{u}_h^1 = \frac{q_{\mrm{res},h}}{|\mathcal{N}_\mrm{s}|}\,\mathbf{1}_{|\mathcal{N}_\mrm{s}|}.
\end{equation}
The remaining components, $ C_{\mathcal{N}_\mrm{l}}\hat{u}_h^1$, are uniquely determined by the load-flow constraint~\eqref{eq:load_match}. Given this initial control trajectory, we obtain the associated state trajectory $\hat{x}_{k}^1$ by solving Simulation Problem~\ref{prob:sim_prob}.

The full method is described in Algorithm \ref{alg:sislp}.

\begin{algorithm}[H] 
\caption{Simulation-Informed Sequential Convex Programming (SI-SCP)} \label{alg:sislp}
\begin{algorithmic}[1] 
\State \textbf{Initialization:} 
Calculate $\hat{u}^{a=1}_{h}$ from \eqref{eq:qres} with $q_{linepack,h}=0 \quad \forall h \in (0,...,N) $. Set $\Delta t^{a=1}_k = 1\, h \quad \forall k = 1, ..., K^{a=1}$. 
With $\hat{u}^{a=1}_{h=0}$ solve~\eqref{eq:steady_state}
to obtain $x^{a=1}_{k=0}$. With $x^{a=1}_{k=0}$ and $\hat{u}^{a=1}_{h}$, solve Simulation Problem~\ref{prob:sim_prob} for $k = 1, ..., K^{a=1}$ to obtain $\hat{x}^{a=1}_{k}$.
\vspace{1ex}

\For{$a = 1, \dots, a_{\mrm{max}}$}
  \State Solve Convex Optimization Problem~\ref{prob:opt_prob} to obtain $u^{\mathrm{opt},a}_{h}$ and $x^{\mathrm{opt},a}_{k}$.
  \State Solve Simulation Problem~\ref{prob:sim_prob} using $u^{\mathrm{opt},a}_{h}$ as boundary values to obtain $x^{\mathrm{sim},a}_{k}$.
  \State Set \textbf{p\_ok} $\gets$ \textbf{True}, \textbf{u\_ok} $\gets$ \textbf{True}.
  \For{$h = 0, \dots, H-1$}
  \vspace{1ex}
    \If{$\| C_p x^{\mathrm{opt},a}_{t_l\in [t_h,t_{h+1})}
        - C_p x^{\mathrm{sim},a}_{t_l\in [t_h,t_{h+1})} \|_{\infty} > \mathrm{tol}_{\mrm{p}}$}
        \vspace{1ex}
        \State $\textbf{p\_ok} \gets \textbf{False}$
        \vspace{1ex}
        \State $\Delta t^{a+1}_{t_k\in [t_h,t_{h+1})} \gets 0.5\,\Delta t^{a}_{t_k\in [t_h,t_{h+1})}$
        \vspace{1ex}
    \EndIf
    \EndFor
    \vspace{1ex}
    \If{$\| u^{\mathrm{opt},a}_h - \hat{u}^{\,a}_h \|_{\infty} > \mathrm{tol}_{\mrm{u}}$}
        \vspace{1ex}
        \State $\textbf{u\_ok} \gets \textbf{False}$
    \EndIf
    
    \If{\textbf{p\_ok} \textbf{and} \textbf{u\_ok}}
        \State \textbf{break}
      \Else
        \vspace{1ex}
        \State $\hat{u}^{\,a+1}_h \gets u^{\mathrm{opt},a}_h$
        \vspace{1ex}
        \State $\hat{x}^{\,a+1}_k \gets x^{\mathrm{sim},a}_k$
        \vspace{1ex}
    \EndIf
\EndFor
\end{algorithmic}
\end{algorithm}

\section{Results and Discussion}\label{sec:results}
In the following section, we solve the optimization problem introduced in Section~\ref{sec:optimization} using the SI-SCP approach from Section~\ref{sec:siscp} to assess whether linepack can significantly contribute to meeting hydrogen storage demand. We consider a full-year horizon on a 40-node test network and compare metrics defined in Section~\ref{sec:metrics} for energy-system scenarios with low, medium, and high hydrogen demand. Additionally, we show that SI-SCP  maintains a high accuracy with regard to the pressure constraints while keeping the problem size small compared to fixed-step approaches.

\subsection{Case Study} \label{sec:case-study}
Constructing a case study for a hydrogen network is challenging as there are no existing test cases or references from large-scale networks already in operation. Since our research question is motivated by energy system model results, we use scenario data from such models, combined with a standard network topology from the natural gas literature.

\subsubsection{Network Topology}
We use the standard test network GasLib-40 with 40 nodes, 45 edges and a total pipeline length of $\qty{1112}{\km}$ from the GasLib library~\cite{Schmidt.2017}. 
Using a natural gas topology for a hydrogen test case is reasonable, as a large part of the future hydrogen network will consist of repurposed natural gas pipelines~\cite{EHB.2023}. The only topology modification is replacing the three compressors with resistors (see~\eqref{eq:valve_equation}), as compressor modeling is beyond our scope.
We assume the following operational constraints of the network: $p_{\mrm{max}} = \qty{70}{\bar}$ and $p_{\mrm{min}} = \qty{20}{\bar}$.

\begin{figure}[h]
\centering
\includegraphics[width=0.7\textwidth]{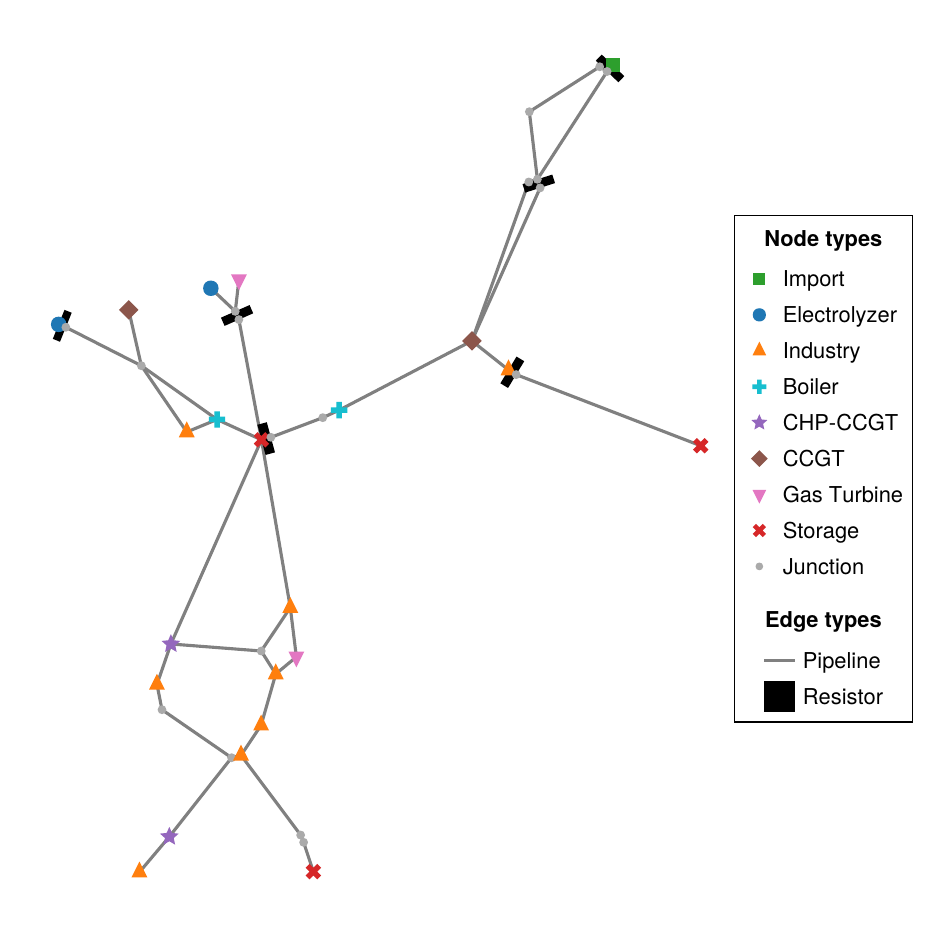}
\caption{GasLib-40 network topology~\cite{Schmidt.2017} with modified node-types and compressor station edges operated in bypass mode (resistor). Nodes with combined cycle gas turbines are referred to as CCGT and nodes with combined heat and power CCGT are referred to as CHP-CCGT.}
\label{fig:gaslib-40}
\end{figure}

\subsubsection{Scenario Data}
To meaningfully apply the proposed methodology, we require scenario data that reflects the future flexibility needs of the hydrogen system. We use hydrogen generation and consumption data from Lux et al., describing a greenhouse-gas-neutral energy system in 2050~\cite{Lux.2022}. The data is available in hourly resolution, aggregated for the whole of Germany. To transfer the data to the GasLib-40 test network, two steps are necessary: distribution to individual network nodes and downscaling.

To distribute the scenario data, we introduce node subtypes, since GasLib-40 distinguishes only supply nodes, demand nodes and junctions. To realistically represent the network's load and flexibility requirements, we account for the different characteristics of producers and consumers: based on~\cite{Lux.2022}, we define two supply-node subtypes and five demand node subtypes. We further repurpose some demand nodes as storage nodes, since external storage is indispensable for our analysis. An overview of the different node subtypes and their representation in the network is provided in Figure \ref{fig:gaslib-40}.

It remains to define what proportion of Germany's total 2050 hydrogen demand is allocated to the GasLib-40 network for our test purposes. We define a reference scenario in which approximately four percent of German demand is to be met by the GasLib-40 network (see Table \ref{tab:scenario_definition}). In addition, we consider a low- and a high-utilization scenario for sensitivity analysis. The
high-utilization scenario is scaled to represent the maximum feasible transport
capacity of the network without the use of compressors. While further increases
in network utilization are theoretically possible, they would necessitate the
inclusion of compressor stations in the optimization model, which is beyond the
scope of this work.

\begin{table}[ht]
\caption{Scenario definition}
    \label{tab:scenario_definition}
    \centering
\begin{tabularx}{\columnwidth}{@{}Xccc@{}}
\toprule
& \textbf{Low demand} & \textbf{Reference} & \textbf{High demand}\\
\midrule
Share of total German demand  & $2.8 \, \%$ & $4.2 \, \%$ & $5.6 \, \%$ \\
Max. storage flow rate per storage & $\qty{102}{\normalcubicmeterpersec}$ & $\qty{153}{\normalcubicmeterpersec}$ & $\qty{204}{\normalcubicmeterpersec}$\\
Number of caverns per storage  & $4$ & $6$ & $8$ \\
Total number of caverns & $12$ & $18$ & $24$\\
\bottomrule
\end{tabularx}
\end{table}

\begin{figure}[h]
\centering
\includegraphics[width=1\textwidth]{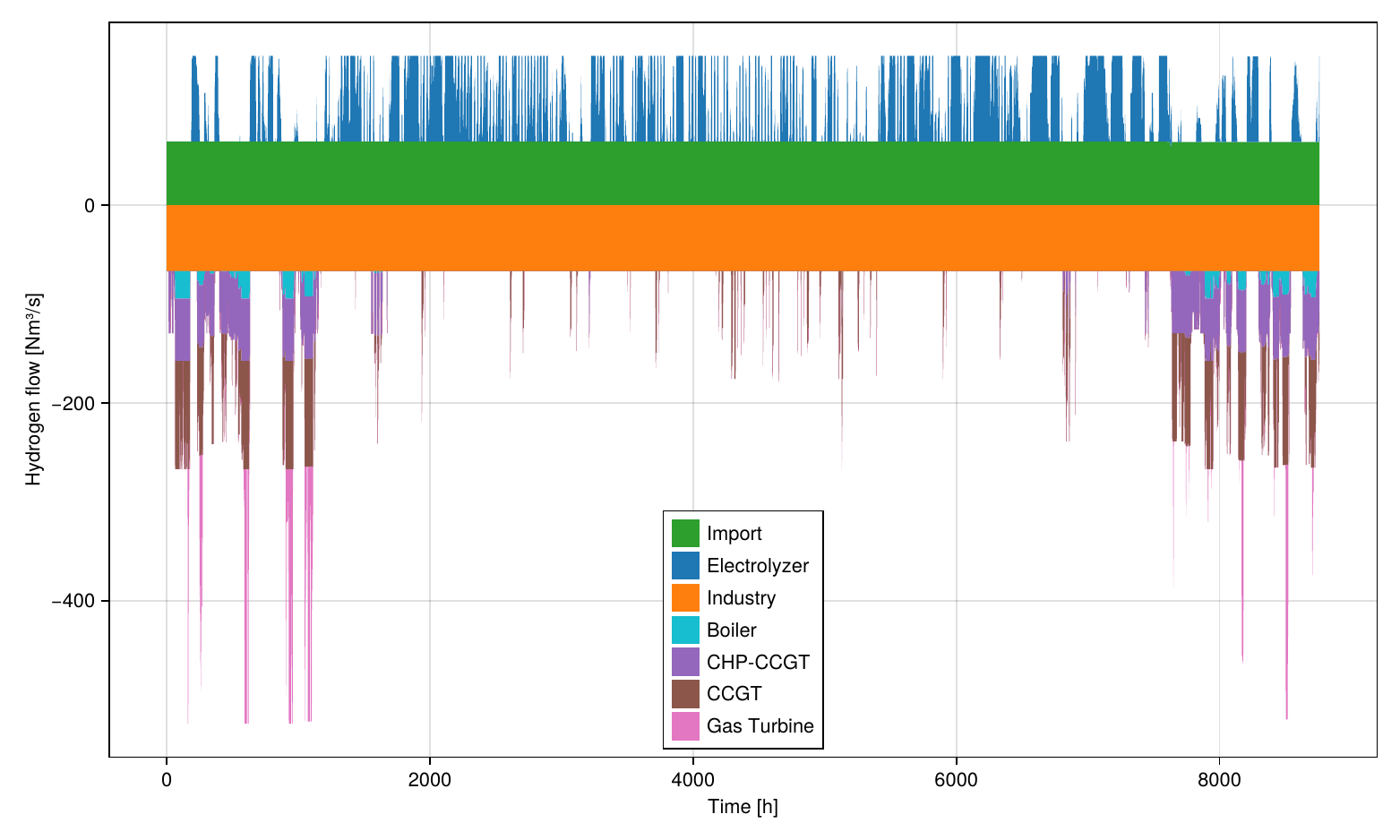}
\caption{Scenario data for the reference scenario, provided from~\cite{Lux.2022} for the whole of Germany and scaled down to the GasLib-40 network. Positive values denote hydrogen supply and negative values denote hydrogen demand. The residual of supply and demand is to be covered either by cavern or by linepack storage.}
\label{fig:ecm_data}
\end{figure}

Figure~\ref{fig:ecm_data} displays the total demand per node subtype excluding storage nodes for the reference scenario, which is distributed evenly across the number of nodes in this subtype. It can be seen that the hydrogen system optimized for the year 2050 is characterized by large peak loads from hydrogen power plants in winter. These peak loads lead both to pronounced pressure transients in the network as well as to pronounced seasonal effects in the storage profile. 

The hourly residual load follows from the difference between supply and demand. In the benchmark scenario, it is covered solely by cavern storage. Assuming a typical cavern size of approximately \qty{45e6}{\normalcubicmeter} (normal cubic meters at \qty{0}{\degreeCelsius} and \qty{1.013}{\bar})~\cite{Elberry.2021}, this results in the number of caverns specified
in Table~\ref{tab:scenario_definition}. In our test case, the caverns are distributed evenly across the three storage locations and operated simultaneously. In the following, we quantify the upper technical limit of linepack storage contribution to residual load and compare the results to the benchmark case.

\subsection{Quantifying the Linepack Storage Capacity}
We employ the proposed SI-SCP approach to quantify the linepack storage capacity. From a full-year optimization, we compute the linepack storage metrics and compare them to other storage options. By comparing the optimized result to the benchmark scenario, we can gain a detailed understanding of the interaction between linepack storage and cavern storage. Applied across different utilization scenarios, the method further allows investigating the influence of hydrogen demand on linepack storage.

\subsubsection{Linepack Storage Metrics}

\begin{figure}[h]
\centering
\includegraphics[width=1\textwidth]{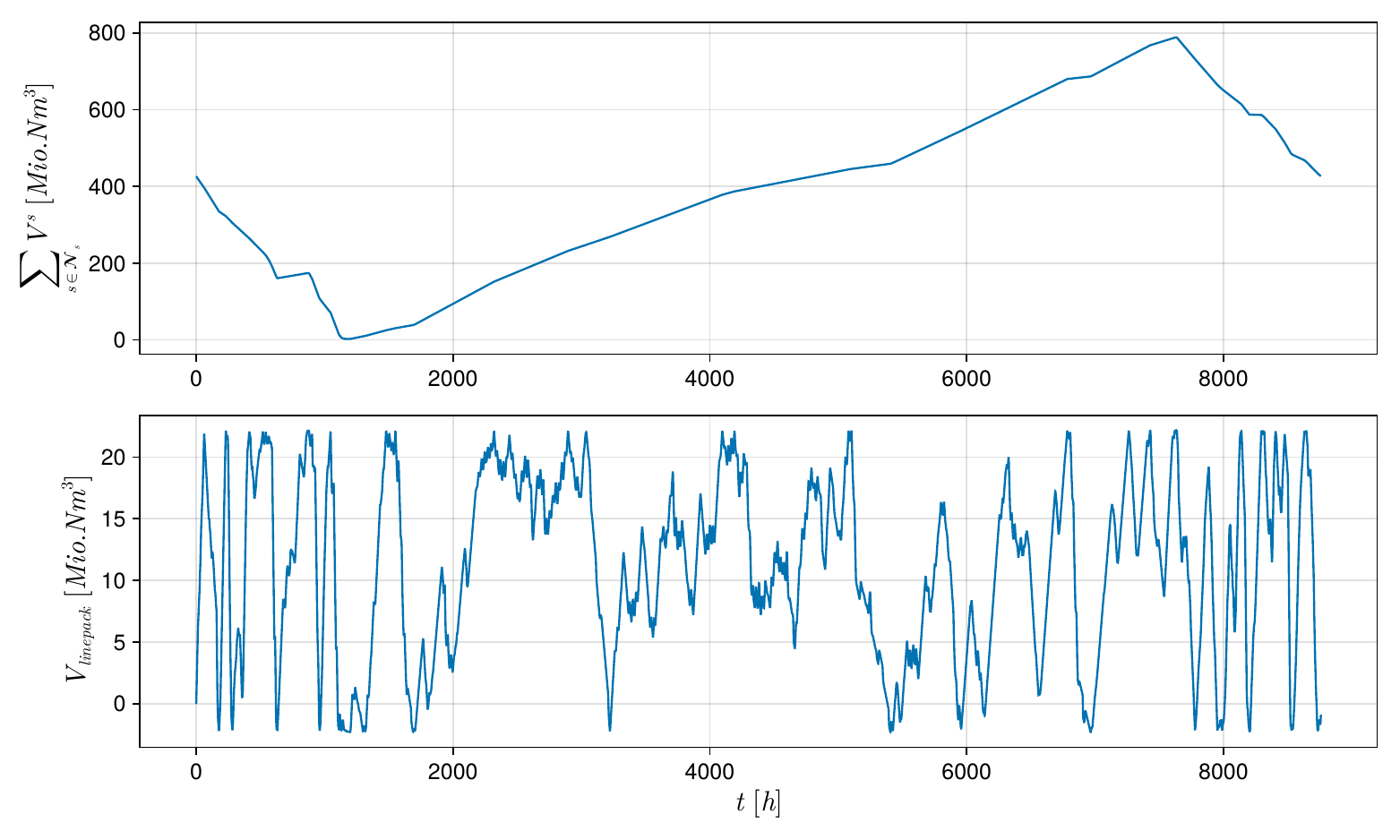}
\caption{Optimized linepack and cavern storage profiles for the reference scenario. The cavern storage facilities operate on a seasonal profile, while linepack storage acts as short-term storage.}
\label{fig:results_agv_comparison}
\end{figure}

The objective of the optimization is to maximize the utilization of linepack storage while minimizing the demand for cavern storage. Using the storage dynamics~\eqref{eq:storage_eq}, both the linepack and cavern storage profiles can be derived from the optimization results (see Figure~\ref{fig:results_agv_comparison}). The linepack storage profile includes negative and positive working gas volumes, where $V_{\mrm{linepack}} = 0$ corresponds to the pressure level used for transport only. The annual profile is characterized by frequent charging and discharging cycles, resulting in $28.79$ full-cycle equivalents. In contrast, the aggregated cavern storage profile shows a nearly strictly seasonal profile with $1.02$ full-cycle equivalents. Thus, a typical division between seasonal storage in the cavern and short- to medium-term storage via linepack is reflected in the optimization results.

The volume capacity of the linepack storage is small compared to the 
total cavern capacity
but large compared to other short- to medium-term storage options. As can be seen in Figure~\ref{fig:results_agv_comparison}, the cavern storage exhibits a volume capacity of $\sum_{s \in \mathcal{N}_s} \Delta V^s \approx \qty{800e6}{\normalcubicmeter}$, whereas the linepack storage reaches only $ \Delta V_{\mrm{linepack}} \approx \qty{24e6}{\normalcubicmeter}$. Since seasonal storage typically requires much larger volume capacities than short-term storage, a comparison of the linepack storage with a typical short-term storage alternative is appropriate. Conventional short- to medium-term gas storage units in natural gas systems are underground pipe storage facilities. A large pipe storage converted to hydrogen would have a maximum working gas volume of approximately $ \Delta V_{\mrm{pipe}} \approx \qty{1e6}{\normalcubicmeter}$~\cite{Elberry.2021}. Thus, the network’s linepack storage can replace approximately 24 pipe storage units in this scenario.

Comparing the annual throughput instead of the volume capacity, it becomes clear that linepack storage contributes significantly to the overall storage demand. Due to the high number of cycles, the linepack storage has an annual throughput of \qty{2.5}{\terawatthoursperyear}. This equates to $89 \, \%$ of the annual cavern storage throughput of \qty{2.8}{\terawatthoursperyear}.

\subsubsection{Interaction Between Linepack and Cavern Storage}
We compare the optimization results with the benchmark scenario, which represents cavern storage operation without any linepack or alternative pipe storage. By analyzing the differences between the benchmark and the maximum linepack scenarios, we can assess the influence of linepack storage on the operation and utilization of cavern storage.

\begin{figure}[h]
\centering
\includegraphics[width=1\textwidth]{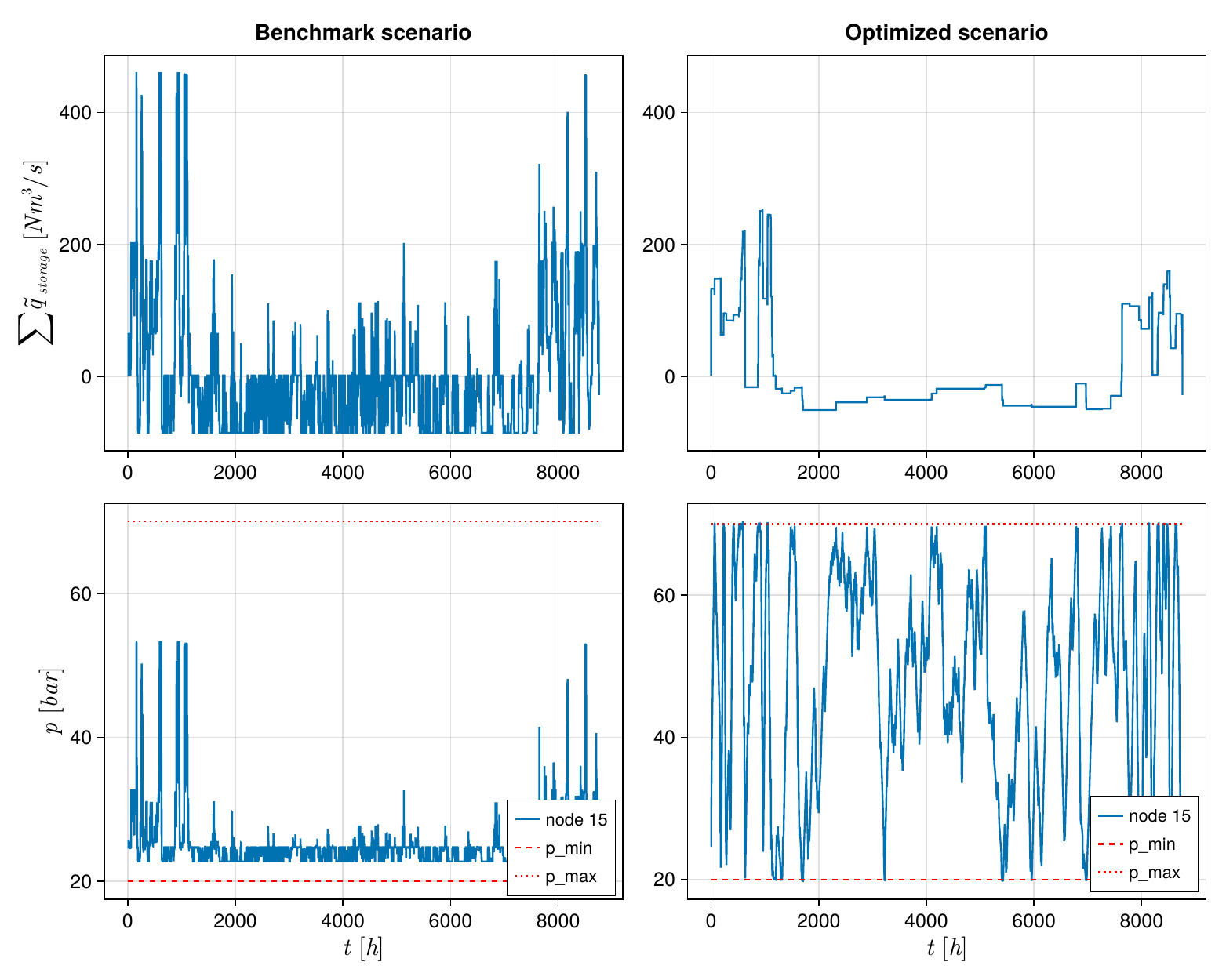}
\caption{Sum of all external storage flows compared to the network pressure at node 15 (one of three storage nodes) for both the benchmark and the optimized scenario. Positive flows represent discharging while negative flows represent charging of the cavern storage. The fluctuations of the storage flow can be substantially reduced through the usage of linepack storage. This requires a very flexible network operation using the full pressure spectrum between maximum and minimum operating pressure.}
\label{fig:results_u_and_p}
\end{figure}

The use of linepack storage significantly reduces both the volume and flow rate capacities required from the cavern storage compared to the benchmark scenario. As shown in Figure~\ref{fig:results_u_and_p}, the maximum charging and discharging flow rate capacities decrease by $41 \,\%$ and $45 \,\%$ relative to the benchmark scenario, respectively. As previously illustrated in Figure~\ref{fig:results_agv_comparison}, the volume capacity is reduced by approximately \qty{24e6}{\normalcubicmeter}, corresponding to a reduction of about $3\,\%$ over all caverns. While the differences in volume and flow rate capacity between the benchmark and optimized scenarios correspond directly to the respective linepack storage capacities, this relationship does not hold for the throughput. The annual throughput of the cavern storage decreases by $24\,\%$ from \qty{3.7}{\terawatthoursperyear} in the benchmark scenario to \qty{2.8}{\terawatthoursperyear} in the optimized scenario, whereas the annual throughput of the linepack storage increases from \qty{0.0}{\terawatthoursperyear} to \qty{2.5}{\terawatthoursperyear}. 

The use of linepack storage also reduces the flexibility requirements of the cavern storage facilities. As shown in Figure~\ref{fig:results_u_and_p}, the cavern flow in the benchmark scenario frequently switches between withdrawal and injection. The number of injection-withdrawal switches decreases from $379$ in the benchmark scenario to $5$ in the maximum linepack scenario, corresponding to a reduction of $98.6\, \%$. However, this reduction in cavern flexibility comes at the cost of increased operational flexibility within the network. While the network pressure in the benchmark scenario remains in a smaller operating range, and especially during the summer months in a very narrow band, the optimized scenario exhibits pronounced pressure fluctuations between the operating limits during the whole year (see Figure~\ref{fig:results_u_and_p}). As both scenarios represent the extreme points of the feasible solution space, it can be expected that real-world operation will involve a trade-off between cavern flexibility and network flexibility.

\subsubsection{Influence of the Hydrogen Demand on the Linepack Storage Potential}
All results shown so far were derived from the reference scenario. To assess how these findings depend on the assumptions made on network utilization, we compare them to the results for the low- and high-demand scenarios. 

The volume capacity of the linepack storage changes only marginally across the three demand scenarios (see Table~\ref{tab:linepack_metrics}). It is slightly lower in the higher demand scenarios due to increased transport-related pressure losses, which reduce the available pressure range for storage. We can calculate the theoretical volume capacity of the network without any pressure losses due to transport as follows
\begin{equation}
    V_{\textrm{no transport}} = \left( p_{\mrm{max}} - p_{\mrm{min}} \right) \frac{V_{\textrm{geom}}}{R_s T z \tilde{\rho}}, 
\end{equation}
with $V_{\textrm{geom}}$ denoting the geometrical volume of the whole network. For the GasLib-40 network, the capacity without transport losses amounts to $V_{\textrm{no transport}} = \qty{24.70e6}{\normalcubicmeter}$. 
This indicates that frictional losses have only a small influence on the volume capacity within the considered scenarios.
Note, however, that this represents the maximum available capacity over the course of the year. During peak load hours, when pressure losses are high, this full capacity is not available.

\begin{table}[ht]
\caption{Linepack storage metrics}
    \label{tab:linepack_metrics}
    \centering
\begin{tabularx}{\columnwidth}{@{}Xcccc@{}}
\toprule
\textbf{Metric}& \textbf{Benchmark - } & \textbf{Optimized - } & \textbf{Optimized - } & \textbf{Optimized - } \\
& \textbf{all scenarios} & \textbf{Low demand} & \textbf{Reference} & \textbf{High demand}\\
\midrule
Max. charge capacity & $0 \, GW$ & $2.9 \, GW$ & $4.2 \, GW$ & $5.0 \, GW$\\
Max. discharge capacity & $0 \, GW$ & $1.6 \, GW$ & $2.3\, GW$ & $3.1 \, GW$\\
Volume capacity & \qty{0e6}{\normalcubicmeter} & \qty{24.6e6}{\normalcubicmeter} & \qty{24.4e6}{\normalcubicmeter} & \qty{24.3e6}{\normalcubicmeter} \\
Full-cycle equivalents & $0$ & $21.2$ & $28.8$ & $35.2$\\
\bottomrule
\end{tabularx}
\end{table}

Although the volume capacity is nearly constant, the operational characteristics of the linepack storage change significantly. Both the maximum charging and discharging flow rate capacities, here reported as power based on the higher heating value, and the number of full-cycle equivalents increase with higher hydrogen demand (see Table~\ref{tab:linepack_metrics}). The rise in full-cycle equivalents indicates that at higher demand levels, the linepack storage is shifted from medium-term storage towards short-term balancing. Consequently, with similar capacity but more cycles, the annual throughput increases substantially.

The absolute metrics provide only part of the picture. To assess the share of the overall residual load that can be covered by linepack storage, they must be related to the cavern storage metrics. While the volume capacity of the linepack storage shrinks slightly at higher demand, the cavern capacity increases by up to $100\,\%$ between low demand and high-demand scenario (see Table~\ref{tab:scenario_definition}). Thus, the ratio between the volume capacities of both kinds of storage decreases strongly. A similar trend is observed for the annual throughput. 
Although the absolute throughput of the linepack storage increases with higher demand, the ratio between linepack storage throughput and cavern storage throughput decreases from $102\,\%$ in the low-demand scenario to $80\,\%$ in the high-demand scenario. This indicates that the contribution that linepack storage can make towards covering the total storage demand decreases as the transport load of the network increases.

Furthermore, we see that cavern storage operation benefits most from linepack storage in the low-demand scenario. Under low demand, cavern storage can operate with an almost purely seasonal profile, exhibiting $1.00$ full-cycle equivalents and $5$ injection–withdrawal switches (see Table~\ref{tab:cavern metrics}).

\begin{table}[ht]
\caption{Cavern storage metrics}
    \label{tab:cavern metrics}
    \centering
\begin{tabularx}{\columnwidth}{@{}Xcccc@{}}
\toprule
\textbf{Metric}& \textbf{Benchmark - } & \textbf{Optimized - } & \textbf{Optimized - } & \textbf{Optimized - } \\
& \textbf{all scenarios} & \textbf{Low demand} & \textbf{Reference} & \textbf{High demand}\\
\midrule
Charge capacity reduction & $0 \, \%$ & $47\, \%$ & $41 \, \%$ & $35 \, \%$\\
Discharge capacity reduction & $0 \, \%$ & $58 \, \%$ & $45\, \%$ & $32 \, \%$\\
Full-cycle equivalents & $1.29$ & $1.00$ & $1.02$ & $1.03$\\
Injection-withdrawal switches & $379$ & $5$ & $5$ & $9$\\
\bottomrule
\end{tabularx}
\end{table}

In addition to the metrics shown, the transient behavior of the network also changes in relation to hydrogen demand. Higher demand levels lead to more pronounced transient pressure peaks. Especially for such scenarios, our SI-SCP algorithm demonstrates significant performance advantages compared to conventional fixed-step-size SCP methods, as shown in the next section.

\subsection{SI-SCP Performance}

Solving the transient hydrogen network optimization problem inherently involves a trade-off between numerical accuracy and computational feasibility. A finer temporal discretization improves the accuracy of the solution but leads to a substantial increase in problem size and computational cost. To balance this trade-off, our SI-SCP algorithm dynamically adapts the time step such that the discretization error remains within a predefined tolerance of $ \mathrm{tol}_{\mrm{p}} = \qty{0.7}{\bar}$ in network pressure.
We choose the tolerance for the linearization error as $1\,\%$
of the maximum individual storage flow. In the Benchmark Scenario,
the entire residual flow is covered by the external storage units,
so that the maximum individual storage flow can be estimated by
distributing $q_\mrm{res}(t)$ equally among all storage nodes:
\begin{equation}
    \mathrm{tol}_{\mrm{u}} = 0.01 \cdot
    \max_{t}\,\frac{|q_\mrm{res}(t)|}{|\mathcal{N}_\mrm{s}|}.
\end{equation}
Additionally, the maximum number of iterations is set to
$a_{\mrm{max}} = 12$.

The challenges faced when solving the optimization problem vary significantly across the different demand scenarios. Table~\ref{tab:SI-SCP_performance} shows
that, while a large number of hours require at least one level of time-step
refinement in all scenarios, the number of hours that require deeper refinement
($\Delta t < \qty{1800}{\s}$) remains very small. In the low-demand scenario,
only $6$ out of $8760$ hours need to be refined beyond a single bisection, the minimum refined time step is $\Delta t = \qty{450}{\s}$, and
the algorithm converges in just $5$ iterations. This fast convergence can be attributed to the lower impact of frictional effects. Since the friction term is the source of nonlinearity in the system, sequential linearization converges faster when friction effects are small. The high accuracy achievable with large time steps can also be explained by the absence of pronounced pressure transients in low-demand conditions. In contrast, the high-demand scenario requires $11$ iterations and a minimum time step of $\qty{28.13}{\s}$ to handle the strong friction-induced nonlinearities and pronounced pressure transients. However, even here only $80$ out of $8760$
optimized hours require refinement beyond $\qty{1800}{\s}$.

This highlights a key advantage of the SI-SCP algorithm: it adapts the time step only where this improves accuracy, which can reduce the problem size significantly. In order to remain below the predefined tolerance, an equidistant time-step SCP approach would result in approximately $1.1 \times 10^6$ time steps in the high-demand scenario. In comparison, our SI-SCP approach requires only $15{,}021$ time steps to achieve the same accuracy, reducing the overall problem size by $98.66\,\%$ (see Table~\ref{tab:SI-SCP_performance}).

\begin{table}[ht]
\caption{Performance of the Simulation-Informed Sequential Convex Programming Algorithm}
    \label{tab:SI-SCP_performance}
    \centering
\begin{tabularx}{\columnwidth}{@{}Xccc@{}}
\toprule
\textbf{Metric} & \textbf{Optimized - } & \textbf{Optimized - } & \textbf{Optimized - } \\
& \textbf{Low demand} & \textbf{Reference} & \textbf{High demand}\\
\midrule
Number of iterations & $5$ & $6$ & $11$\\
Hours with refined $\Delta t$ & $3661$ & $4745$ & $5377$\\
Hours with $\Delta t < 1800 \;\textrm{s} $  & $6$ & $23$ & $80$\\
Minimum $\Delta t$ & \qty{450.00}{\s} & \qty{112.50}{\s}& \qty{28.13}{\s} \\
Number of time steps $K$ & $12{,}441$& $13{,}745$ & $15{,}021$\\
Problem size reduction compared to equidistant time steps & $82.25\,\%$ & $95.10 \, \%$ & $98.66\,\%$\\
\bottomrule
\end{tabularx}
\end{table}

\section{Conclusion}\label{sec:conclusion}
The aim of this paper was to develop a numerical method that can be used to reliably and efficiently quantify the potential contribution of linepack storage to the overall hydrogen storage demand. Moreover, we sought a better understanding of the interaction between linepack storage and cavern storage in hydrogen systems. 

To achieve this goal, we introduced a new class of transient optimization problems characterized by two key features that distinguish it from the state of the art in natural gas research: a different objective function and an extended optimization horizon. To keep the problem size tractable, we proposed a novel SI-SCP approach that uses adaptive temporal resolution to ensure high accuracy with a minimum number of time steps. 

Investigating a case study with a 40-node network, we solved the optimization problem for a full-year horizon in order to capture seasonal effects, which play an important role for future hydrogen storage.
Building on the optimization results, we evaluated metrics for the linepack storage such as flow rate capacity and volume capacity allowing direct comparison with cavern and pipe storage facilities. 
Our results show that the optimized linepack storage achieves an annual throughput comparable to the cavern storage, but through a fundamentally different operating pattern: a small volume capacity is combined with a high number of cycles, whereas the cavern storage operates with high volume capacity and a mainly seasonal profile.
The use of linepack storage reduces the flow rate and volume capacity required from cavern storage. More importantly, it absorbs nearly all short- and medium-term fluctuations, substantially lowering the flexibility requirements on the caverns.
The sensitivity analysis further shows that the ratio of linepack to cavern storage capacity decreases with increasing network utilization. The relative contribution of the network to the overall storage demand is therefore largest in low-utilization scenarios.

The proposed SI-SCP algorithm reduces the problem size by 82--98\,\% compared to standard SCP with a fixed time step. This gain stems from the local nature of the resolution requirements: while pronounced pressure transients demand time steps as short as \qty{28}{\s} to achieve the required pressure accuracy, fewer than 1\,\% of all hours require steps below \qty{30}{\min} in any scenario.

We emphasize that our study addresses the technical linepack storage
potential. To this end, we exploited the full flexibility of the network,
enforcing only compliance with technical constraints, which in our case are
the maximum and minimum operating pressures. The choice of this admissible
pressure range influences the technical linepack storage potential and must
be assessed on a case-by-case basis for real hydrogen networks. The technical
linepack storage capacity can be regarded as an upper-bound
estimate that does not account for economic or regulatory limitations.
In future work, we aim to extend this approach by integrating economic assessments to quantify the cost of linepack storage and benchmark it against alternative storage technologies. Ultimately, this will support the design of cost-efficient and technically viable hydrogen storage portfolios for a future hydrogen economy.

\section*{CRediT authorship contribution statement}
\textbf{Ulrike Herrmann:} Conceptualization, Methodology, Software, Validation, Writing - Original Draft, Writing - Review \& Editing, Visualization. \textbf{Hans Würfel:} Methodology, Software, Writing - Review \& Editing. \textbf{Albertus Johannes Malan:} Conceptualization, Validation, Writing - Review \& Editing. \textbf{Anton Plietzsch:} Conceptualization, Writing - Review \& Editing, Supervision. \textbf{Johannes Schiffer:} Conceptualization, Writing - Review \& Editing, Supervision.

\section*{Declaration of competing interests}
The authors declare that they have no known competing financial interests or personal relationships that could have appeared to influence the work reported in this paper. 

\section*{Declaration of generative AI and AI-assisted technologies in the manuscript preparation process}
During the preparation of this work the authors used ChatGPT and Claude in order to help with software development and to improve language and readability of the manuscript. After using these tools, the authors reviewed and edited the content as needed and take full responsibility for the content of the published article.

\section*{Acknowledgments}
The authors gratefully acknowledge Benjamin Lux for kindly providing the scenario data used in this study. Furthermore, the authors gratefully acknowledge the financial support by the German Federal Ministry of Education and Research, as part of the public research projects TransHyDE (03HY201S) and EIZ (03SF0693E), as well as the German Federal Ministry for Economic Affairs and Climate Action, as part of the OpPoDyn-Project (03EI1071A).

\appendix
\numberwithin{equation}{section}
\numberwithin{figure}{section}
\numberwithin{table}{section}

\section{Three-node example network}\label{sec:appendix_A}
\begin{figure}[h]
\centering
\includegraphics[width=1\textwidth]{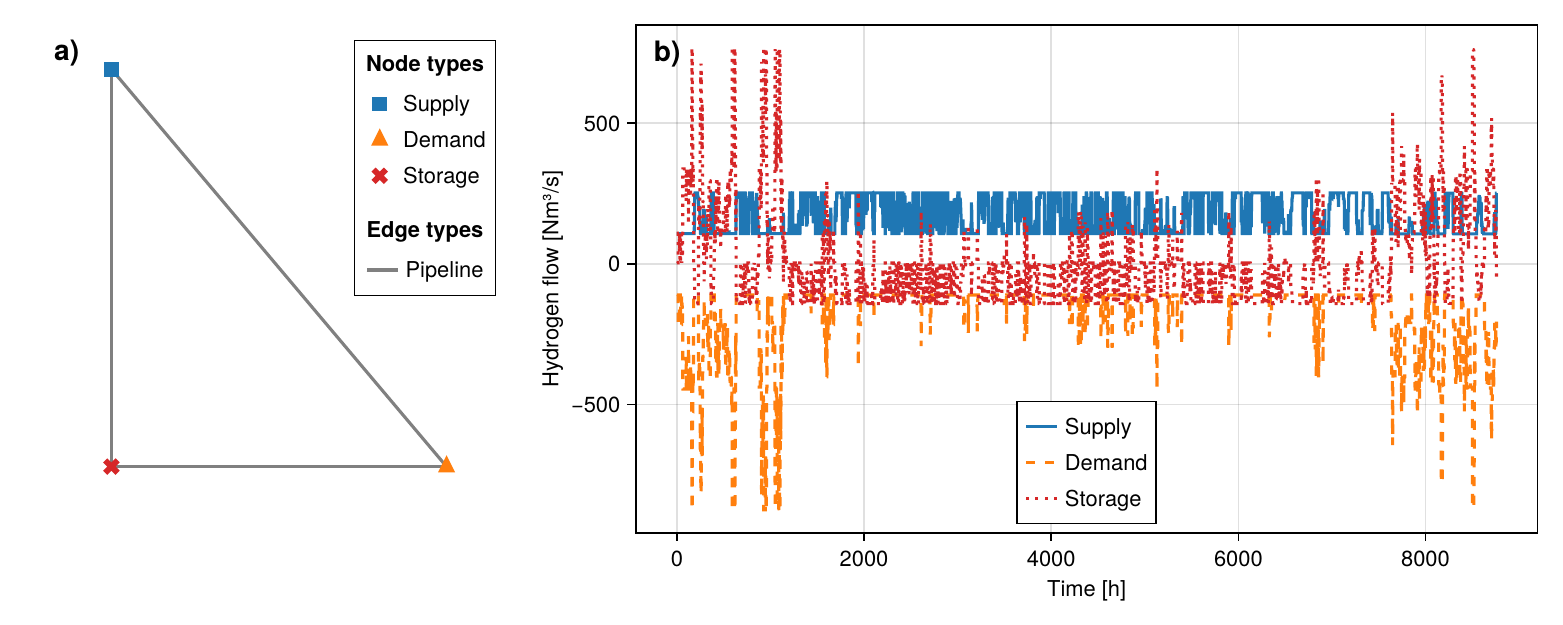}
\caption{a) Three-node example network topology. b) Scenario data for the three-node example network scaled to cover $7\,\%$ of the total German hydrogen demand in~\cite{Lux.2022}.}
\label{fig:three-node-example}
\end{figure}

The simple meshed three-node network from~\cite{Malan.2023} is shown in Figure~\ref{fig:three-node-example} a) and is used to demonstrate the results of Algorithm~\ref{alg:slp}. The scenario data is obtained from~\cite{Lux.2022}. We do not distinguish different supply and demand node subtypes. Instead, we aggregate all supply and demand profiles respectively and downscale the profiles to cover $7.00\,\%$ of the total German supply and demand profiles. The storage flow in Figure~\ref{fig:three-node-example} b) is shown for the benchmark scenario, i.e. $q_\mrm{res}(t) = -q_\mrm{stor}(t)$.

\section{SCP algorithm with fixed time step}\label{sec:appendix_B}

Following the basic SCP idea presented e.g. in~\cite{OMalley.2021, Mhanna.2022, Raheli.2025}, we formulate a simple SCP algorithm with fixed time step in Algorithm \ref{alg:slp}.

\begin{algorithm}[H]  
\caption{Sequential Convex Programming (SCP)} \label{alg:slp}
\begin{algorithmic}[1] 
\State \textbf{Initialization:} 
Calculate $\hat{u}_{h}$ from \eqref{eq:warmstart}. With $\hat{u}_{0}$ solve~\eqref{eq:steady_state}
to obtain $x_{0}$. With $x_{0}$ and $\hat{u}_{h}$, solve Simulation Problem \ref{prob:sim_prob} for $k = 1, ..., K$ to obtain $\hat{x}_{k}$.
\vspace{1ex}

\For{$a=1,\dots,a_{\mrm{max}}$}
  \State Solve Convex Optimization Problem~\ref{prob:opt_prob} to obtain $u^{a}_{h}$ and $x^{a}_{k}$.
    \If{ $|| u^{a}_h - \hat{u}^{a}_h||_{\infty} < \mathrm{tol}_\mrm{u}$ }
      \State \textbf{break}
      \Else 
      \State $\hat{u}^{a+1}_h \gets u^{a}_h$
      \State $\hat{x}^{a+1}_k \gets x^{a}_k$
    \EndIf
\EndFor
\end{algorithmic}
\end{algorithm}



\bibliographystyle{IEEEtran}
\bibliography{LinepackPaper}

\end{document}